\documentclass[twocolumn]{aastex62}

\usepackage{amsmath,amssymb,amstext}
\usepackage{natbib}
\newcommand{\um}{$\mu$m}

\shorttitle{Lensed Red QSO}
\shortauthors{Glikman et al.}

\begin{document}

\title{A Highly Magnified Gravitationally Lensed Red QSO at $z=2.5$ with a Significant Flux Ratio Anomaly}

\author[0000-0003-0489-3750]{Eilat Glikman}
\affiliation{Department of Physics, Middlebury College, Middlebury, VT 05753, USA }

\author[0000-0003-4561-4017]{Cristian E. Rusu}
\affiliation{Subaru Fellow, Subaru Telescope, National Astronomical Observatory of Japan, 650 N Aohoku Pl, Hilo, HI 96720, USA}
\affiliation{National Astronomical Observatory of Japan, 2-21-1 Osawa, Mitaka, Tokyo 181-0015, Japan}

\author[0000-0002-2580-6321]{Geoff C.-F. Chen}
\affiliation{Department of Physics, University of California, Davis, CA 95616, USA}

\author{James Hung-Hsu Chan}
\affiliation{Institute of Physics, Laboratory of Astrophysique, École Polytechnique Fédérale de Lausanne (EPFL), Observatoire de Sauverny, 1290, Versoix, Switzerland}

\author[0000-0002-2231-6861]{Cristiana Spingola}
\affiliation{Dipartimento di Fisica e Astronomia, Universit\`{a} degli Studi di Bologna, Via Gobetti 93/2, I$-$40129 Bologna, Italy}
\affiliation{INAF $-$ Istituto di Radioastronomia, Via Gobetti 101, I$-$40129, Bologna, Italy}

\author[0000-0002-8999-9636]{Hannah Stacey}
\affiliation{Max Planck Institute for Astrophysics, Karl-Schwarzschild Str 1, D-85748 Garching bei München, Germany}

\author{John McKean}
\affiliation{Kapteyn Astronomical Institute, University of Groningen, PO Box 800, NL-9700 AV Groningen, the Netherlands}

\author[0000-0001-5538-5903]{Ciprian T. Berghea}
\affiliation{U.S. Naval Observatory, 3450 Massachusetts Avenue NW, Washington, DC 20392, USA}

\author[0000-0002-0603-3087]{S.~G. Djorgovski}
\affiliation{California Institute of Technology, 1200 E. California Boulevard, Pasadena, CA 91125, USA}

\author[0000-0002-3168-0139]{Matthew J.~Graham}
\affiliation{California Institute of Technology, 1200 E. California Boulevard, Pasadena, CA 91125, USA}

\author[0000-0003-2686-9241]{Daniel Stern}
\affiliation{Jet Propulsion Laboratory, California Institute of Technology, Pasadena, CA 91109, USA}

\author[0000-0001-6746-9936]{Tanya Urrutia} 
\affiliation{Leibniz Institut f\"{u}r Astrophysik, An der Sternwarte 16, D-14482 Potsdam, Germany}

\author[0000-0002-3032-1783]{Mark Lacy}
\affiliation{National Radio Astronomy Observatory, Charlottesville, VA 22903, USA}

\author[0000-0002-4902-8077]{Nathan J. Secrest} 
\affiliation{U.S. Naval Observatory, 3450 Massachusetts Avenue NW, Washington, DC 20392, USA}

\author[0000-0002-7893-1054]{John M.~O'Meara}
\affiliation{Department of Physics, Saint Michael's College, One Winooski Park, Colchester, VT, 05439, USA}
\affiliation{W.M. Keck Observatory 65-1120 Mamalahoa Highway, Kamuela, HI 96743, USA}

\begin{abstract}
We present the discovery of a gravitationally lensed dust-reddened QSO at $z=2.517$, identified in a survey for QSOs by infrared selection. {\em Hubble Space Telescope} imaging reveals a quadruply lensed system in a cusp configuration, with a maximum image separation of $\sim1.8\arcsec$. We find that compared to the central image of the cusp, the neighboring brightest image is anomalous by a factor of $\sim 7-10$, which is the largest flux anomaly measured to date in a lensed QSO. 
Incorporating high-resolution Jansky Very Large Array radio imaging and sub-mm imaging with the Atacama Large (sub-)Millimetre Array, we conclude that a low-mass perturber is the most likely explanation for the anomaly.
The optical through near-infrared spectrum reveals that the QSO is moderately reddened with $E(B-V)~\simeq~0.7-0.9$.
We see an upturn in the ultraviolet spectrum due to $\sim~1\%$ of the intrinsic emission being leaked back into the line of sight, which suggests that the reddening is intrinsic and not due to the lens.   
The QSO may have an Eddington ratio as high as $L/L_{\rm Edd}~\approx~0.2$. 
Consistent with previous red QSO samples, this source exhibits outflows in its spectrum as well as morphological properties suggestive of it being in a merger-driven transitional phase. 
We find a host-galaxy stellar mass of $\log{M_\star/M_\odot}=11.4$, which is higher than the local $M_{BH}$ vs. $M_\star$ relation, but consistent with other high redshift QSOs. 
When de-magnified, this QSO is at the knee of the luminosity function, allowing for the detailed study of a more typical moderate-luminosity infrared-selected QSO at high redshift. 
\end{abstract}


\section{Introduction} \label{sec:intro}

Models of galaxy evolution that invoke major mergers \citep[e.g.,][]{Sanders88a,DiMatteo05,Hopkins05} have been highly successful at incorporating the growth of supermassive black holes (SMBH) in galactic nuclei and explaining various scaling relations between the two, such as the $M-\sigma$ relation \citep{Ferrarese00,Gebhardt00}. These models predict a phase during the merger process in which the growing SMBH is enshrouded by dust.  And, while at its peak luminosity, this active galactic nucleus (AGN, or, the more luminous QSO\footnote{In this work, we adopt the canonical nomenclature that distinguishes quasars – radio-detected luminous AGN whose radio emission is essential to their selection – from QSOs – the overall class of luminous AGN.}) is heavily obscured and thus elusive to most AGN and QSO surveying techniques, especially at visible wavelengths. Recent work in the near-infrared has revealed a population of QSOs with moderate amounts of dust extinction that appear to be transitioning from a heavily dust-enshrouded phase to a typical, unobscured QSO \citep[e.g.,][]{Glikman12,Banerji12,Brusa15}.  

Combining near-infrared and radio data has proven to be a very effective method for finding quasars in this transitional state \citep{Glikman04,Glikman07,Urrutia09,Glikman12,Glikman13}. These efforts have resulted in a sample of $\gtrsim120$ dust-reddened quasars from the combined FIRST+2MASS surveys (F2M) with reddenings in the range $0.1 < E_{B-V} < 1.5$. F2M red quasars are found to be predominantly driven by major mergers \citep{Urrutia08,Glikman15}, are accreting at very high rates \citep[$L/L_{\rm Edd} \simeq 0.69$;][]{Kim15} and exhibit broad absorption lines associated with outflows and feedback \citep{Urrutia09}. 
These properties are consistent with buried quasars expelling their dusty shrouds in an evolutionary phase predicted by merger-driven coevolution models. 

Among the sources in the F2M sample, two gravitationally lensed systems were found.  
F2M~J0134$-$0931 is a radio-loud red quasar at $z=2.216$ that is lensed into at least 5 images, possibly by two galaxies at $z=0.7645$ \citep{Gregg02,Winn02,Hall02a}. 
This scenario \citep{Keeton03} proposes that the lenses are both spiral galaxies, which may then also be responsible for the reddening.  
F2M~J1004+1229, at $z=2.65$, is a rare low-ionization broad absorption line quasar (LoBAL) that includes strong absorption from metastable \ion{Fe}{2} \citep[FeLoBAL;][]{Becker97}.  The location of the reddening in this system is unclear \citep{Lacy02}.

Based on distinct color differences between optically-selected lensed QSOs and those selected in the radio or infrared, \citet{Malhotra97} suggest that reddening by dust in the lensing galaxy is biasing surveys for lensed QSOs, underestimating their numbers.  Alternatively, if the reddening of the lensed QSOs is intrinsic, then their larger presence suggests that the population of red quasars may be significantly underestimated. 

Both of the lensed F2M quasars were selected requiring a radio detection, a wavelength that is largely insensitive to dust reddening, which may have made them easier to find. However, since radio-loud and radio-intermediate quasars make up only $\sim10$\% of the overall quasar population \citep{Ivezic02}, the radio restriction also limited the sample to a rarer class of quasars. 
In this paper, we report the discovery of a quadruply lensed radio-quiet red QSO discovered in a search for red QSOs using {\em WISE} color selection and no radio criterion. 

Throughout this work we quote magnitudes on the AB system, unless explicitly stated otherwise. 
When computing luminosities and any other cosmology-dependent quantities, we use the $\Lambda$CDM concordance cosmology: $H_0=70$ km s$^{-1}$ Mpc$^{-1}$, $\Omega_M=0.30$, and $\Omega_\Lambda=0.70$.

\section{Discovery and Observations} \label{sec:discovery}

\subsection{Selection}\label{sec:selection}

We recently constructed a sample of radio-quiet dust-reddened QSOs selected by their infrared colors in {\em WISE} and 2MASS (W2M), applying well-established color cuts in {\em WISE} color space \citep{Lacy04,Stern05,Donley12,Stern12,Assef13,Assef18,Glikman18} and the infrared-to-optical KX color space \citep{Warren00}.  
Our survey covers $\sim 2000$ deg$^2$ with a relatively shallow near-infrared flux limit ($K<16.7$) and has resulted in 37 newly-identified red QSOs \citep{Glikman22}.  Among the sources was W2M~J104222.11+164115.3\footnote{The source name is shortened to W2M~J1042+1641 hereafter.}, whose infrared luminosity, based on {\em WISE} photometry, $L_{IR} \simeq 10^{14} L_\odot$, was more luminous than any other known radio-quiet QSO and implied extreme properties suggestive of gravitational lensing. 

The source is undetected in FIRST, implying that its 20 cm flux density is below 1 mJy.  
There are also multi-epoch {\em Swift}/X-ray Telescope (XRT) observations of this object revealing moderately high absorption ($N_H \sim 10^{23}$ cm$^{-2}$) and exhibiting variability \citep{Matsuoka18}.
Table \ref{tab:photometrysurveys} lists the broadband magnitudes of this source from the surveys used in its discovery. 




\begin{deluxetable}{cc|cc}



\tablewidth{0pt}

\tablecaption{Integrated Photometry of W2M~J1042+1641 from available surveys
\label{tab:photometrysurveys}}

\tablenum{1}

\tablehead{\colhead{Band} & \colhead{AB Mag} & \colhead{Band} & \colhead{AB Mag}  } 

\startdata
$u$\tablenotemark{a}  &    20.93 $\pm$ 0.10 & $H$\tablenotemark{b}  &    16.87 $\pm$ 0.13 \\
$g$\tablenotemark{a}  &    20.40 $\pm$ 0.03 & $K_s$\tablenotemark{b}  &  15.87 $\pm$ 0.06 \\
$r$\tablenotemark{a}  &    20.26 $\pm$ 0.03 & $W1$\tablenotemark{c}  &   15.52 $\pm$ 0.02 \\
$i$\tablenotemark{a}  &    20.04 $\pm$ 0.04 & $W2$\tablenotemark{c}  &   14.97 $\pm$ 0.02 \\
$z$\tablenotemark{a}  &    18.99 $\pm$ 0.05 & $W3$\tablenotemark{c}  &   13.00 $\pm$ 0.02 \\
$J$\tablenotemark{b}  &    17.84 $\pm$ 0.17 & $W4$\tablenotemark{c}  &   11.84 $\pm$ 0.04 \\
\enddata
\tablenotetext{a}{SDSS Model magnitudes from de Vaucouleurs profile fitting.}
\tablenotetext{b}{2MASS magnitudes.}
\tablenotetext{c}{AllWISE magnitudes.}

\end{deluxetable}

\subsection{Initial Spectroscopy} \label{sec:spec1}

W2M~J1042+1641 was observed with the MODS1B Spectrograph on the Large Binocular Telescope (LBT) observatory for 1200 sec, with the red and blue arms simultaneously, with a 0\farcs6-wide slit on UT 2013 March 14, covering the wavelength range $3300 - 10100$\AA.  After removing the CCD signatures ({\tt modsCCDred}), spectral extraction, wavelength and flux calibration, and telluric correction were done with the IRAF {\tt apall} task. 

On UT 2013 March 19, we observed the source with the SpeX spectrograph \citep{Rayner03} on the NASA Infrared Telescope Facility (IRTF) for 32 minutes using an 0\farcs8-wide slit covering a wavelength range of $0.808 - 2.415$\um.  
The seeing was 1\arcsec\ and sky conditions were clear.  An A0V star was observed within an airmass difference of 0.1 immediately after the object spectrum was obtained to correct for telluric absorption.  The data were reduced using the Spextool software \citep{Cushing04} and the telluric correction was conducted following \citet{Vacca03}.  

\begin{figure*}[ht!]
\plotone{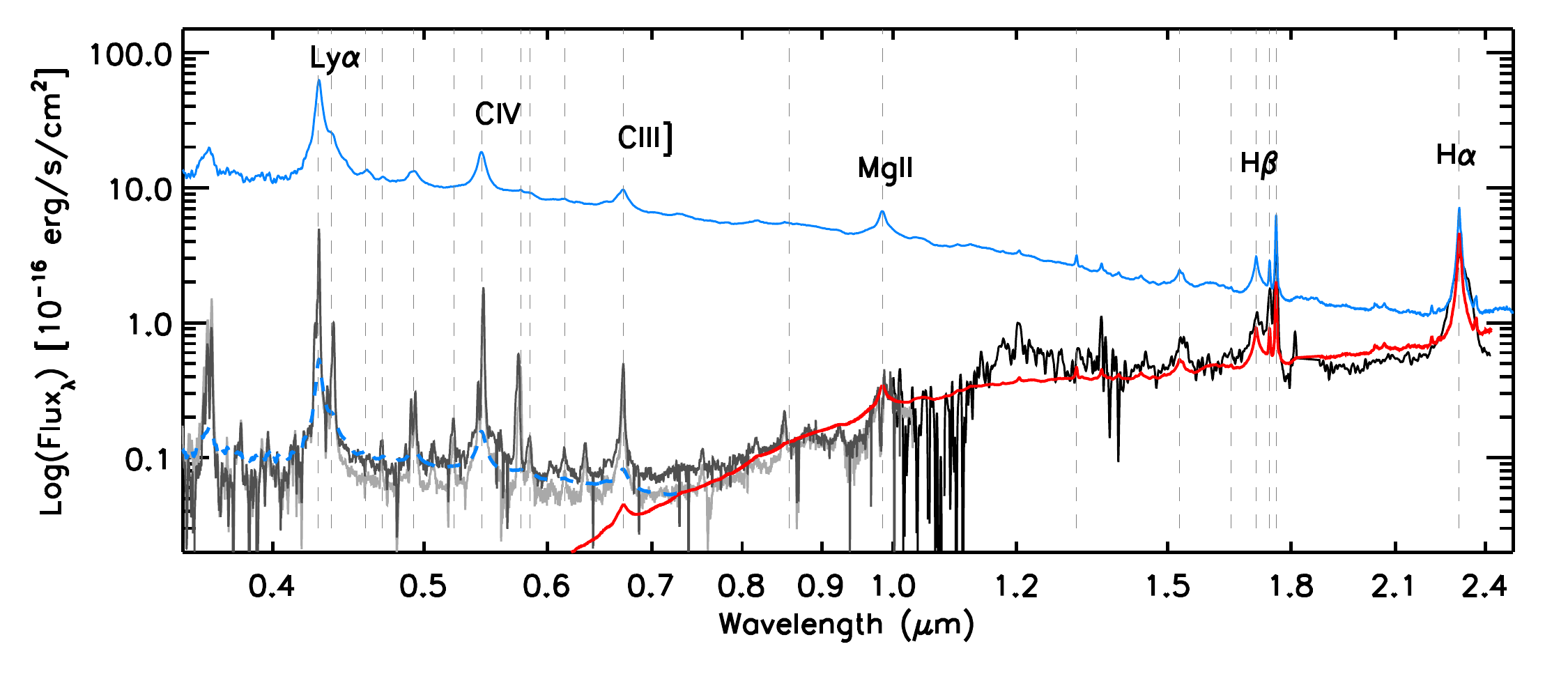}
\caption{Optical-through-near-infrared spectrum of W2M~J1042+1641, plotted on logarithmic wavelength and flux axes. The black line at $\lambda > 1 \mu$m is the near-infrared spectrum, showing broad Balmer line emission shifted to $z=2.517$.  The dark grey line is the LBT optical spectrum and the light grey line is the LRIS spectrum (\S \ref{sec:lris}), showing the consistency between the two spectra taken five years apart (observed frame).
Vertical dashed lines mark the locations of strong emission lines seen in the spectrum. 
The red line is the best-fit QSO template (shown in blue) reddened by $E(B-V)=0.68$ to the LBT spectrum combined with the near-infrared spectrum. In both cases, the rest-frame UV part of the spectrum deviates from the reddened template, implying that the obscuring dust is shielding most of the central region but allows $\sim 0.8\%$ of the intrinsic emission to enter our line of sight (blue dashed line). 
}\label{fig:spec1} 
\end{figure*}

Figure \ref{fig:spec1} shows the combined optical-through-infrared spectrum of W2M~J1042+1641.  The near-infrared spectrum shows strong broad H$\alpha$ and H$\beta$ plus the narrow [\ion{O}{3}] doublet at $\lambda\lambda4959,5007$, while the optical spectrum shows narrow emission lines in permitted as well as forbidden species, securing a QSO redshift identification of $z=2.517$. 
The blue curve represents an unreddened QSO spectrum, made out of the UV composite QSO template of \citet{Telfer02} combined with the optical-to-near-infrared composite spectrum from \citet{Glikman06}, illustrating the large amount of UV light lost. We fit this curve to the spectrum following the technique outlined in \citet{Glikman07} and find that a suitable fit can only be achieved if the rest-frame UV emission below 2275\AA\ ($\lambda_{\rm obs} < 8000$\AA) is ignored.  This best fit is achieved with a QSO template reddened by $E(B-V)=0.68$ (corresponding to a $A_V = 5.4$ mag in the QSO rest frame; red line). 

The excess UV flux, blue-ward of $\sim 8000$\AA, can be explained if the dust were placed close to the AGN, between the broad- and narrow-line-emitting regions. This interpretation is also consistent with a model for the UV spectrum of Mrk 231 (a nearby, dusty, luminous QSO in a merger) suggested by \citet{Veilleux13} in which the broad-line region is reddened by a dusty and patchy outflowing gas. This model predicts a small ``leakage'' fraction of a few percent, which is consistent with a similar degree of leakage seen in the X-ray spectra of other red quasars \citep{Glikman17}.  
A similar conclusion was reached by \citet{Assef16} for a hot Dust Obscured Galaxy (DOG) that displayed blue excess in its spectral energy distribution (SED). The authors arrive at leaked intrinsic QSO light through a patchy obscuring medium, or by reflection, as the best explanation.

We plot in Figure \ref{fig:spec1} the QSO template scaled to 0.8\% of the intrinsic spectrum (with a dashed blue line) and find that it fits well the spectral shape.  We note that the UV emission lines have a higher equivalent width, but are narrower than the template, similar to `extremely red' QSOs in SDSS studied by \citet{Hamann17}.
These arguments lead us to conclude that the dust is local to the lensed QSO and that the QSO is not reddened by the lens. 

\subsection{Hubble Imaging} \label{sec:hst}

We obtained {\em Hubble Space Telescope} ({\em HST}) imaging of W2M~J1042+1641 with the WFC3/IR camera in Cycle 24 as part of a program to study the host galaxies of W2M red QSOs.  We used the F160W and F125W filters, which were chosen to straddle the 4000\AA\ break.  We observed the source over two visits, UT 2017 February 26 and UT 2017 May 7, covering both filters in a single orbit observation per visit.  We observed our sources in MULTIACCUM mode using the STEP100 sampling, which is designed to provide a broad dynamic range while avoiding saturation. We performed a 4-point box dither pattern with 400 (224) sec at each position for the F160W (F125W) filter. We reduced the images using the {\tt DrizzlePac} software package to a final pixel scale of 0\farcs06 pixel$^{-1}$. 

\begin{figure}[ht!]
\begin{center}
\includegraphics[angle=0,scale=0.6]{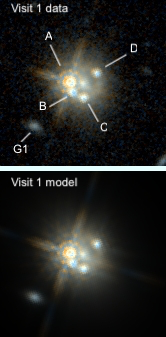}
\includegraphics[angle=0,scale=0.6]{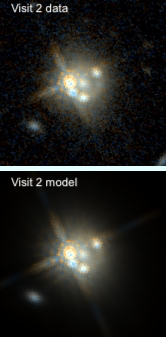}
\caption{{\em HST} WFC3/IR F125W and F160W color-combined images of W2M~J1042+1641 over two visits, along with output from a morphological analysis with {\tt hostlens}. The top row shows the observed, drizzled image. The second row shows the best fit model consisting of four PSFs, S\'{e}rsic profiles for the lensing galaxy (located in between the four PSFs) and G1, and the Einstein ring image of the source host. 
The modeled components are labeled and referenced in the text.    
The images are oriented with north pointing up and east to the left, although the two visits were each taken rotated by 47.2$^\circ$ with respect to each other. The scale is $10\arcsec \times 10\arcsec$.
{\em Left --} Visit 1, UT 2017 February 26. {\em Right --} Visit 2, UT 2017 May 7.
}\label{fig:hst1}
\end{center}
\end{figure}

The top row of Figure \ref{fig:hst1} shows the reduced, color-combined images for the two {\em HST} visits, with the first visit shown on the left and the second visit shown on the right. The image reveals four point sources surrounding an extended-appearing source at the center, in a geometry suggestive of quadruply lensed system with a cusp configuration.
The bottom row of Figure \ref{fig:hst1} shows the results of our profile fits.
The {\bf top row labels} the four lensed components A, B, C, D, in decreasing order of flux as well as nearby galaxy G1, which we also modeled {\bf (see \S \ref{sec:morphmodeling})}. 

\begin{figure}[ht!]
\begin{center}
\includegraphics[angle=0,scale=1]{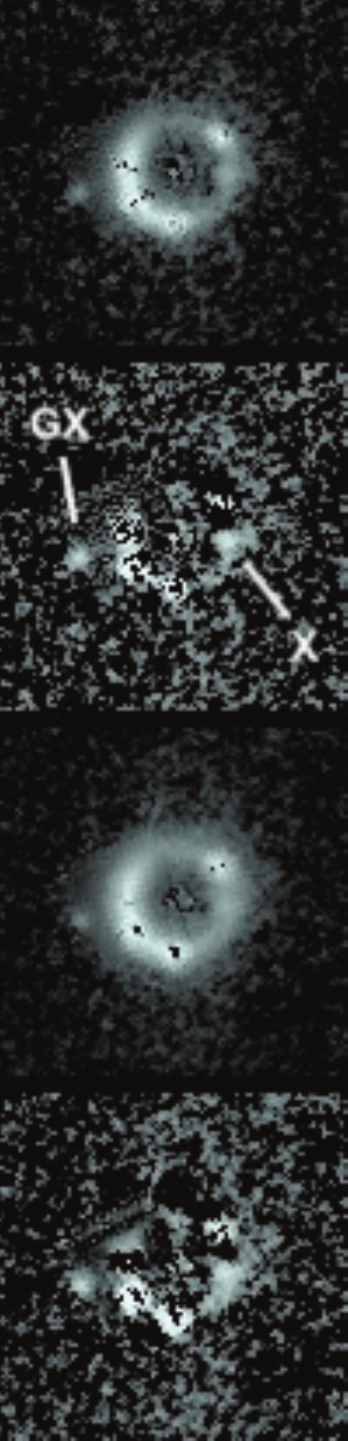}
\includegraphics[angle=0,scale=1]{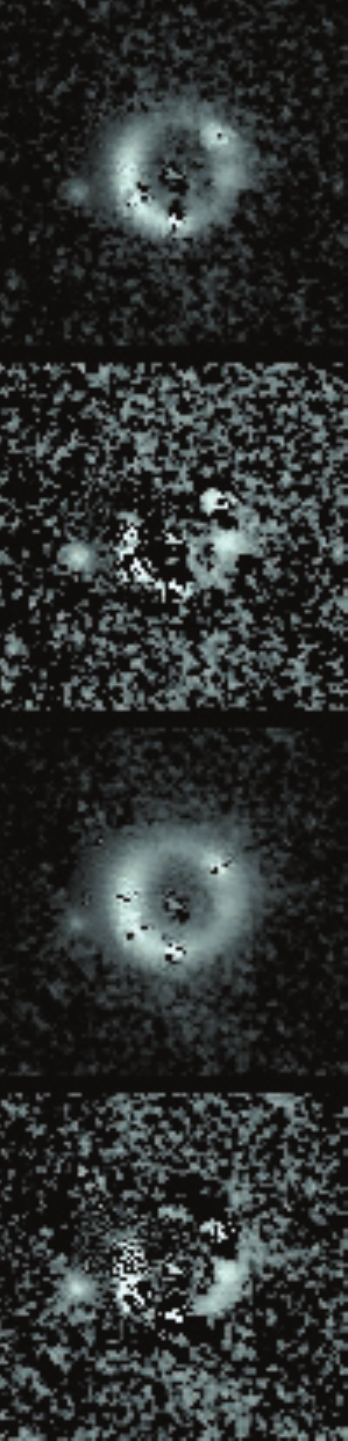}
\caption{{\em HST} Residual images of W2M~J1042+1641 over two visits, after subtracting the best-fit morphological models determined by {\tt hostlens} shown in the bottom panel of Figure \ref{fig:hst1}.   
The top row shows the F125W image with the PSFs and S\'{e}rsic profiles subtracted, revealing a clear Einstein ring made up of the lensed QSO's host galaxy light. The second row shows the full residuals of the same fitted model for the F125W image. The third and fourth rows show the same but for F160W.
The labeled components are the nearby object (GX), which we consider as a galaxy perturber associated to the lensing galaxy in our modeling, and a source that is likely associated with the QSO host galaxy (X).    
The images are scaled to $6\arcsec \times 6\arcsec$ and orientated to match those Figure \ref{fig:hst1}. 
{\em Left --} Visit 1, UT 2017 February 26. {\em Right --} Visit 2, UT 2017 May 7. }
\label{fig:hst1_resid}
\end{center}
\end{figure}

Figure \ref{fig:hst1_resid} shows the image of the Einstein ring after the QSO components (A, B, C, D), modeled lens galaxy, as well as G1 have been subtracted. 
The F125W image is shown in the top row and the second row shows the residuals once the $\sim0.9\arcsec$-radius Einstein ring is also subtracted. Rows three and four show the same but for F160W.
In Tables \ref{tab:photometry}, \ref{tab:astrometry}, and \ref{tab:morphology} we list the resulting photometry, relative astrometry of each component, and the morphological parameters of the lensing galaxy and its companion, respectively. 
 All the {\it HST} data used in this paper can be found in MAST: \dataset[10.17909/rw0m-7191]{http://dx.doi.org/10.17909/rw0m-7191}.
 
\subsection{Follow-up Keck Spectroscopy Along Multiple Position Angles}  \label{sec:lris}

On UT 2018 March 19, we obtained a followup spectrum with the LRIS spectrograph on the Keck I telescope, orienting the slit along different position angles (PAs), aiming to disentangle the emission from the different components.  We placed a slit along the parallactic angle (79$^\circ$) centered on the brightest component for two 600 s exposures.  Another two 600 s exposures were taken with the slit placed along the A, B, C  components, at a position angle of 41.9$^\circ$.  Finally, a fifth 600 s exposure was performed with a position angle of 128.2$^\circ$ along components B and D including the lensing galaxy with the intention of identifying the redshift of the lens.  Although the seeing was $\sim1\arcsec$, precluding our ability to cleanly separate the different components along the position axis of the slit, the 2D spectrum is clearly extended beyond the width of a PSF. Specifically, the data taken at PA $= 128.2^\circ$ shows two clear lensed AGN components separated by $\sim 1\farcs6$, however we detect no obvious signal from the lens itself.  

The combined LRIS spectrum is shown in light grey in Figure \ref{fig:spec1}.    
The best-fit reddened QSO template to the combined LRIS plus near-infrared spectrum, considering only $\lambda > 8000$\AA, finds $E(B-V)=0.73$ (corresponding to a $A_V = 5.8$ mag in the QSO rest frame).  The LRIS and LBT spectra are remarkably similar, suggesting that not much has changed in this source between the two spectroscopic epochs, 5 years apart in the observed frame, or 1.4 years in the rest frame. 

\subsubsection{Lens Redshift} \label{sec:lensz}

\begin{figure}[ht!]
\plotone{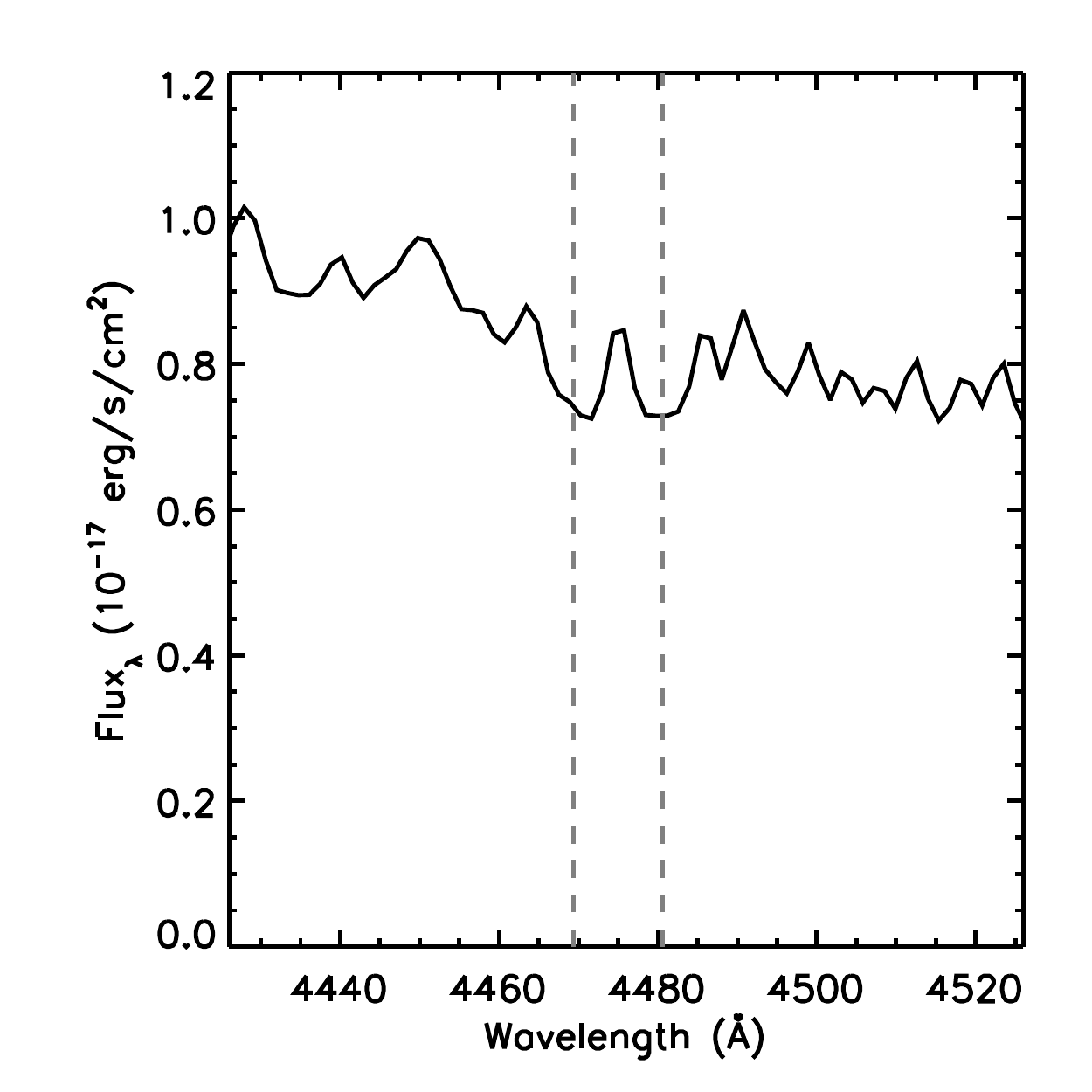}
\caption{We identify a \ion{Mg}{2} $\lambda\lambda$2796,2803 absorption feature, marked by dashed vertical grey lines, at a redshift of $z=0.5985$.}\label{fig:mg2}
\end{figure}

We identify absorption consistent with \ion{Mg}{2} $\lambda\lambda$2796,2803 at $z=0.5985$, Figure \ref{fig:mg2},  
which is in excellent agreement with the photometric redshift expected from the color of the lensing galaxy, and we thus adopt this as a tentative redshift for the lensing galaxy.\footnote{We used the \texttt{mag2mag} routine from \citet{Auger09}, available at \url{https://github.com/tcollett/LensPop/tree/master/stellarpop/}, to check that for a \citet{Coleman80} E/S0 galaxy template, redshifted to $z=0.599$, 19.19 mag in F160W corresponds to 19.58 mag in F125W, in excellent agreement with the lensing galaxy photometry we measured in Table \ref{tab:photometry}.}

\subsection{JVLA Radio follow-up}\label{sec:radio}

We obtained radio data of W2M J1042+1641 with the Jansky Very Large Array (VLA; ID: 19A-430, PI: N. Secrest) on 2019-08-26 (epoch 1) and 2019-09-28 (epoch 2) at C-band (6cm) in A configuration, as part of a program to follow-up quadruply lensed quasars with sensitive VLA observations.
The two observations use the new 3-bit sampler, which gives a 4 GHz bandwidth, divided in 32 spectral windows (spws), with 128 MHz bandwidth and 32 channels each using dual polarization. At the beginning of both epochs we observed J1331+3030 for the amplitude and bandpass calibration, while J1051+2119 was the phase-reference calibrator. 
The scans on the target were $\sim10$ min each, which were interleaved by $\sim2$ min scans on the phase-reference calibrator. 
The total exposure time for each epoch was 90 mins. 
The data were reduced with the Common Astronomy Software Application package \citep[CASA;][]{mcmullin07} following the standard calibration procedures \citep[e.g.,][]{spingola20a,spingola20b}. We detected and CLEANed the target using natural weights. The signal-to-noise ratio was too low to perform self-calibration.

Only emission corresponding to images A and B is detected and resolved. The contour maps of two epochs are shown in Figure \ref{fig:radio}.  
The beam size in the first epoch is $0\farcs300 \times 0\farcs223$ at a position angle of $52\fdg581$ (east of north) and the total integrated flux density of images A and B were of {\bf $51\pm8$ $\mu$Jy and $13\pm7$ $\mu$Jy}, respectively; the off-source rms noise is of 6.7 $\mu$Jy beam$^{-1}$. 
In the second epoch, the beam size is $0\farcs293 \times 0\farcs204$ at a position angle of $-49\fdg727$ degrees  (east of north). Here, the flux densities increased by 50\%, being {\bf $82\pm9$ $\mu$Jy and $19\pm5$ $\mu$Jy} (images A and B, respectively). The off-source rms noise level was 4.6 $\mu$Jy beam$^{-1}$. 
We consider the uncertainty due to calibration (estimated using the scatter on the amplitude gains) to be on the order of 10\%. 
Finally, a 2D Gaussian fit using the task {\sc imfit} to the second epoch CLEANed image (because of the more robust detection) found that both lensed images are consistent with a point source. We discuss the impact of this data on our analysis in \S \ref{sec:magnif}. 
 
 \subsection{ALMA mm imaging}\label{sec:alma}
 
W2M J1042+1641 was observed with the Atacama Large (sub-)Millimetre Array (ALMA) on 2019-10-30 under project code 2019.1.00964.S (PI: Stacey). The target data were correlated in four spectral windows centred on 247, 249, 262 and 264 GHz, each with 2 GHz bandwidth and 128 channels. One of the spectral windows covers the redshifted rest-frequency of a CO line, not reported here. J1058+0133  was observed as a flux and spectral bandpass calibrator. J1045+1735 was used to correct time-dependent phase variations. The total integration time on-target was 5 minutes. The data were calibrated using the ALMA pipeline within CASA and the data were inspected to confirm the quality of the calibration. The continuum-only spectral channels were imaged and deconvolved using a Briggs weighting of the visibility data, resulting in a synthesised beam of $0.64\arcsec\times0.57\arcsec$ and an rms noise of 66 $\mu$Jy. The task {\sc imfit} within CASA was used to fit a PSF to each lensed image. No significant residuals remain after fitting, suggesting that the lensed images are not resolved. 
 
\begin{figure}
\begin{center}
\includegraphics[angle=0,scale=0.35]{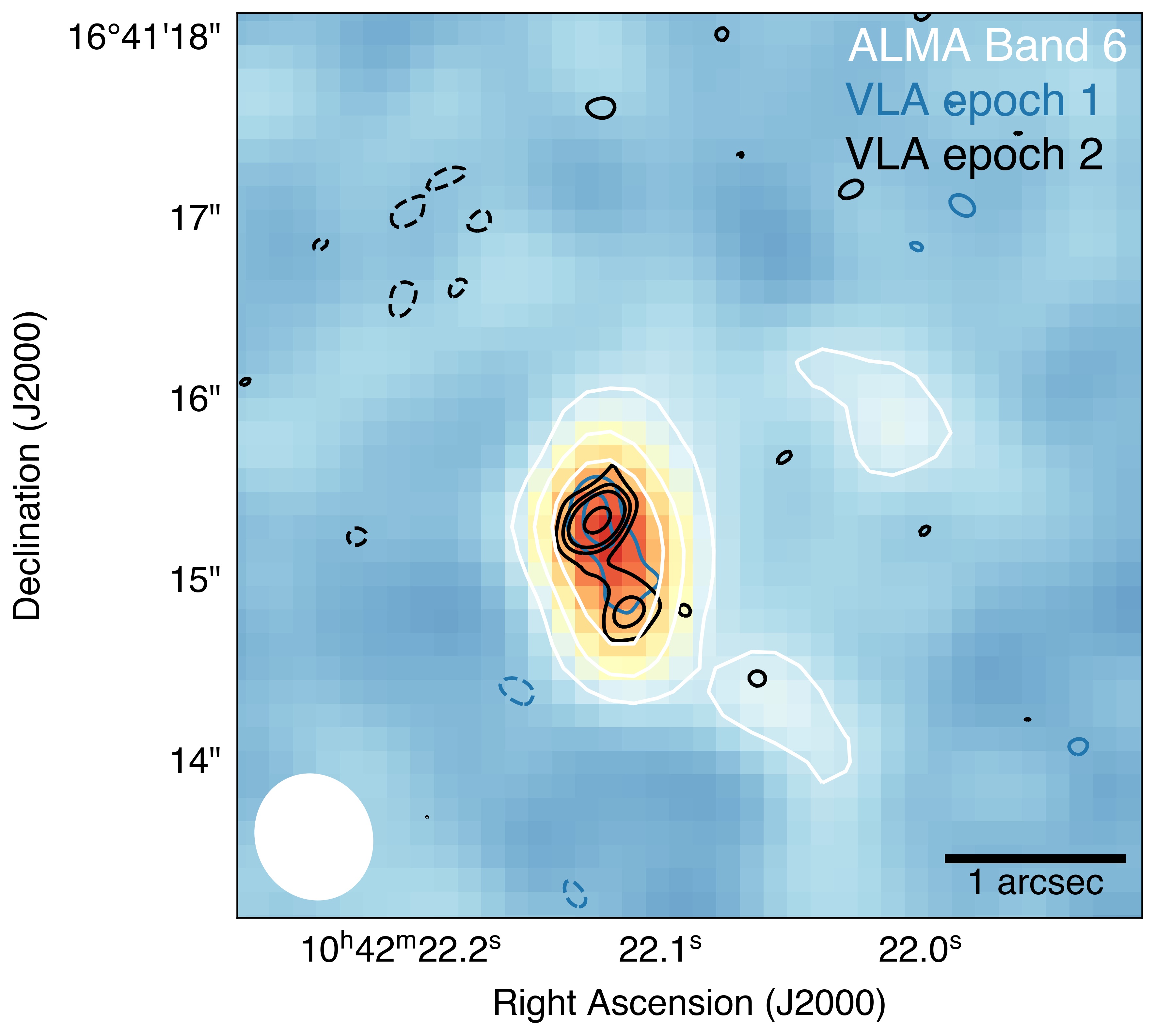}
\caption{Radio (VLA) and sub-mm (ALMA) follow-up imaging. The blue image and black contours indicate the VLA emission at the first and second epochs, respectively. The white contours indicate the ALMA Band 6 emission. The contours are drawn at ($-3$, 3, 6, 9, 18, 36, 72, and 144) times the off-source noise of each map, which is 6.7 $\mu$Jy beam$^{-1}$, 4.6 $\mu$Jy beam$^{-1}$ and 62  $\mu$Jy beam$^{-1}$ for the VLA epoch 1, VLA epoch 2 and ALMA observations, respectively. The restoring  Gaussian beam of the ALMA observations is shown in white the bottom left corner 
and is $0\farcs64 \times 0\farcs57$ with a position angle of $23^\circ$.}\label{fig:radio}
\end{center}
\end{figure}

We overplot in Figure \ref{fig:radio} flux density measurements of the rest-frame 330 $\mu$m continuum, obtained with ALMA. All four QSO images are detected, with flux densities A $=835\pm66 \mu$Jy, B $=519\pm66 \mu$Jy, C $=296\pm66 \mu$Jy and D $=281\pm66 \mu$Jy, consistent with thermal dust emission \citep[e.g.,][]{stacey18}. 
We find no significant evidence of misalignment between the radio (VLA) and sub-mm (ALMA) emissions, indicating a cospatial origin. 
Table \ref{tab:radio_mm} lists the positions and flux densities of the radio and mm sources. 




\begin{deluxetable*}{cllc}




\tablecaption{Radio and mm source positions and flux densities \label{tab:radio_mm}}

\tablenum{2}

\tablehead{\colhead{Source} & \colhead{RA} & \colhead{Dec} & \colhead{Flux density } \\ 
\colhead{} & \colhead{(J2000)} & \colhead{(J2000)} & \colhead{($\mu$Jy)} } 
\startdata
\multicolumn{4}{c}{JVLA Epoch 1} \\
\hline
A    &     10:42:22.1245$\pm$0.0013  &  +16:41:15.3127$\pm$0.0254 &  51$\pm$5   \\
B    &     10:42:22.111$\pm$0.022      &  +16:41:14.942$\pm$0.169     &  13$\pm$1 \\
\hline
\multicolumn{4}{c}{JVLA Epoch 2} \\
\hline
A    &     10:42:22.1252$\pm$0.0005  &  +16:41:15.3371$\pm$0.0073 &  82$\pm$8 \\
B    &     10:42:22.1113$\pm$0.0021  &  +16:41:14.830$\pm$0.0294   &  19$\pm$2 \\
\hline
\multicolumn{4}{c}{ALMA} \\
\hline
A    &     10:42:22.1208$\pm$0.0013  &  +16:41:15.3465$\pm$0.0225  &  835$\pm$66 \\
B    &     10:42:22.1183$\pm$0.0021  &  +16:41:14.8356$\pm$0.0362  &  519$\pm$66 \\
C    &     10:42:22.0441$\pm$0.0037  &  +16:41:14.3003$\pm$0.0634  &  296$\pm$66  \\
D    &     10:42:22.0204$\pm$0.0039  &  +16:41:16.1130$\pm$0.0669  &  281$\pm$66 \\
\enddata

\tablecomments{\bf The ALMA absolute flux calibration uncertainty is $\sim10\%$, https://arc.iram.fr/documents/cycle7/ALMA\_Cycle7\_Technical\_Handbook.pdf }


\end{deluxetable*}

\section{Modeling of the system} \label{sec:morphmodeling}

The Einstein ring shown in Figure \ref{fig:hst1_resid} is relatively bright, and therefore any morphological fitting of the system that does not account for it may bias the quantities of interest: the relative astrometry and photometry of each light source, as well as the morphology of the lensing galaxy. We therefore chose to fit the system using \texttt{hostlens} \citep{Rusu16}, which incorporates, along with the other morphological components, a model of the Einstein ring as an analytical S\'{e}rsic profile concentric with the QSO light, lensed through a lensing mass model and convolved with the point spread function (PSF). 
We focused on a $16\arcsec \times 19\arcsec$ cutout around the system starting with a newly constructed PSF \citep[following the method described in][{\bf which involved combining a few dozen bright stars in each {\em HST} filter.}]{Glikman15} for each filter. 
The images from the two visits were taken at different angles, making their combined-image PSF difficult to model; we therefore model these independently, although some of the parameters are treated as coupled, as we describe in the next section.

\subsection{Coupled light and lens modeling} \label{sec:coupledmodeling}

There are three reasons why a naive, direct modeling of this system with \texttt{hostlens} would be sub-optimal: (1) employing \texttt{hostlens} with a single smooth lens mass model (and without multiple mass substructures whose parameters are highly degenerate) cannot account for the flux ratio anomalies found in this system (see \S  \ref{sec:lensing}); This is because, for a given {\bf mass model}, \texttt{hostlens} does not allow one to arbitrarily change the flux ratios of the images, which are determined by the relative position of the source with respect to the lens.  (2) The PSFs, although carefully constructed in \S  \ref{sec:morphmodeling}, were found to produce significant residuals when used to fit the QSO images, particularly the bright image A. (3) \texttt{hostlens} can only model a single input image cutout at a time. But, given that we have images from two {\em HST} visits in two filters, it is desirable to model the system using the joint information from the cutouts of all available data, in order to better constrain some of the morphological parameters we derive. 

To tackle the issues above, we wrote a custom wrapper code around \texttt{hostlens}, which uses an iterative approach:
\begin{enumerate}

\item We first run \texttt{hostlens} without a lensing model, fitting the four QSO images as point sources characterized by their positions and fluxes; the light of the lensing galaxy G and of the nearby galaxy G1 are fitted with one S\'{e}rsic profile each, convolved with our original PSFs, in each of the two visits and two filters (4 cutouts). During the fitting, our wrapper performs the parameter optimization by minimizing the sum of the quality-of-fit $\chi^2$ reported by \texttt{hostlens} for the 4 cutouts, using the Nelder-Mead algorithm \citep{Gao12}. 
The sky pedestals, relative astrometry, and fluxes of each light component are optimized independently for each cutout\footnote{The reason we do not enforce that the relative positions of each components match between cutouts is that the uncertainties on these positions (especially that of the position of the lensing galaxy) has a dominant effect on the best-fit lens mass models we derive in the next steps, and we found that Markov Chain Monte Carlo (MCMC) approaches to determine this uncertainty can significantly underestimate it. We therefore prefer to use the scatter in relative astrometry between the 4 cutouts as a measure of uncertainty.}. 
For each cutout, we constrain the orientations of the two S\'{e}rsic profile, accounting for the different rotation angle of the two visits. Their ellipticity, effective radius, and S\'{e}rsic index are constrained to matching values in the two visits, but not in the two filters. 

\item We fit a shared lens mass model (details are provided in \S \ref{sec:lensing}) with \texttt{glafic} \citep{Oguri10} using the relative astrometry we derived in the previous step. \texttt{glafic} solves the lens equation and computes the lensed point-source images using an adaptive grid algorithm; it compares these positions to the observed ones via $\chi^2$ minimization, in order to optimize the mass model. We then repeat the optimization from the previous step, but also fitting for the extended QSO host galaxy, responsible for the Einstein Ring, with a circular S\'{e}rsic profile (we found that allowing for ellipticity did not improve the fit), lensed through the lens mass model. We fixed the S\'{e}rsic index to the fiducial value for an early-type galaxy, $n=4$ \citep{devaucouleurs48}, as we found that this parameter would otherwise diverge to large values ($n>10$) without producing visually improved residuals or modifying significantly the photometry of the other components. The fluxes of the host are fitted independently to the four cutouts\footnote{While we do not expect the flux of the host to vary between the two visits, we nevertheless obtained a better fit (fewer residuals) by allowing the host flux to be a free parameter between visits. The final difference is up to $\sim0.2$ mag (see Table \ref{tab:photometry}).}, while the effective radius can vary between the filters but not between visits. 

\item At this point, we found that if we optimize for all light components at the same time, the parameters of the S\'{e}rsic profiles of the lens and QSO host galaxies are affected by the significant residuals at the location of the QSO image. We therefore hold fix the best-fit models of the QSO images and mask them using circular masks of $\sim0.3\arcsec$ in radius\footnote{We also mask the luminous blob on top of the Einstein ring, which we describe in Appendix \ref{sec:host}.}. Next, we optimize the parameters of the S\'{e}rsic profiles and also the astrometry and photometry of object GX (see Figure \ref{fig:hst1_resid}), which we model as a point source\footnote{We attempted to fit GX with a Sersic profile, but we measured a vanishingly small effective radius $\lesssim0.01\arcsec$. We therefore consider this component to be unresolved.}. We then hold fix the parameters of the profiles we just fitted at this step, we remove the masks and fit again for the QSO images, to allow them to adjust in response to the profiles mentioned above. 

\item We follow the approach in \citet{Chen16}, developed for the analysis of gravitational lenses observed with adaptive optics (AO), where the PSFs are a priori unknown and must be derived from the data. This approach improves the PSF of the four cutouts under the assumption that the PSF should not vary among the QSO images. 

\item We now proceed in an iterative fashion, where we first refit the parameters of the shared lens mass model with \texttt{glafic} using the improved relative astrometry\footnote{We checked that the difference between the positions predicted by the lens model and the ones actually measured stay within a fraction of a pixel size for the four cutouts.} from step 3, and repeating steps 3 and 4. We do this in 30 steps, where the PSF correction box size is increased from 7 to 35 pixels on a side, by two pixels every second step. The gradual increase is adopted in order to improve the convergence of the PSF correction, by preventing it from being dominated by noise outside the core of the PSF. Figure \ref{fig:hst1_resid} shows the residuals after the final iteration.\footnote{The shape of the PSF correction box may be seen around image A. As this image is much brighter than the other ones, it has more weight in the improved PSF.} Following the iterations, the $\chi^2$ is much improved, whether we measure it by first masking the pixels corresponding to the QSO images cores or not.
\end{enumerate}

Once we have inferred the best-fit profile parameters as described above, we determine the corresponding uncertainties by combining 5 independent MCMC chains of 15,000 - 30,000 steps. We ensure their convergence by monitoring the change in the parameter values over time and removing the ``burn-in`` steps. Due to our modeling approach, we need to run MCMC separately, first for the QSO images, and then for the other profiles.\footnote{An alternative modeling approach which can model all profiles at the same time is presented, e.g., in \citet{Wong17}. It works by rescaling the weights of the pixels corresponding to large residuals.} While our reconstructed PSF is superior to the original one, it is not perfect, as shot noise is present in the core of the bright point sources. If we integrate the residual flux (positive or negative) in the pixels corresponding to the core of each of the point sources, it is not exactly zero for a given point source. Therefore, for the photometry of the QSO images and the lensing galaxy reported in Table \ref{tab:photometry}, we add to the uncertainties the contribution of this residual flux.




\begin{deluxetable*}{cccccccccc}



\tablewidth{0pt}

\tablecaption{Photometry of W2M~J1042+1641}

\tablenum{3}

\tablehead{
\colhead{Filter} & \colhead{A (mag)} & \colhead{B (mag)} & \colhead{C (mag)} & \colhead{D (mag)} & \colhead{G (mag)} & \colhead{G1 (mag)} & \colhead{GX (mag)} & \colhead{S (mag)}} 

\startdata
F125W (1; w/ GX) & 18.26$\pm$0.001 & 20.48$\pm$0.03 & 21.13$\pm$0.02 & 21.84$\pm$0.02 & 19.57$\pm$0.01 & 23.29$\pm$0.03 & 25.43$\pm$0.05  & 23.30$\pm$0.02 \\ 
F125W (1; w/o GX) & 18.24$\pm$0.0005 & 20.47$\pm$0.03 & 21.09$\pm$0.03 & 21.78$\pm$0.02 & 19.53$\pm$0.01 & 23.29$\pm$0.02 & 25.45$\pm$0.05  & 23.47$\pm$0.02 \\ 
F125W (2; w/ GX) & 18.26$\pm$0.002 & 20.46$\pm$0.004 & 20.92$\pm$0.01 & 21.95$\pm$0.03 & 19.60$\pm$0.01 & 23.23$\pm$0.03 & 25.43$\pm$0.05 & 23.53$\pm$0.02 \\ 
F125W (2; w/o GX) & 18.25$\pm$0.001 & 20.46$\pm$0.01 & 20.90$\pm$0.01 & 21.91$\pm$0.03 & 19.56$\pm$0.01 & 23.23$\pm$0.02 & 25.30$\pm$0.05 & 23.66$\pm$0.02 \\ 
F160W (1; w/ GX) & 17.62$\pm$0.0003 & 20.22$\pm$0.04 & 20.66$\pm$0.05 & 21.49$\pm$0.02 & 19.19$\pm$0.01 & 23.03$\pm$0.02 & 25.22$\pm$0.04 & 22.34$\pm$0.02 \\ 
F160W (1; w/o GX) & 17.60$\pm$0.0004 & 20.22$\pm$0.03 & 20.60$\pm$0.06 & 21.41$\pm$0.02 & 19.18$\pm$0.004 & 23.03$\pm$0.01 & 25.44$\pm$0.05 & 22.20$\pm$0.01 \\ 
F160W (2; w/ GX) & 17.72$\pm$0.001 & 20.24$\pm$0.02 & 20.43$\pm$0.01 & 21.37$\pm$0.004 & 19.16$\pm$0.005 & 22.97$\pm$0.02 & 25.10$\pm$0.04 & 22.38$\pm$0.02 \\ 
F160W (2; w/o GX) & 17.70$\pm$0.001 & 20.22$\pm$0.01 & 20.39$\pm$0.005 & 21.31$\pm$0.01 & 19.19$\pm$0.004 & 22.98$\pm$0.01 & 25.42$\pm$0.05 & 22.23$\pm$0.01 \\ 
\enddata


\tablecomments{Photometry has been measured with \texttt{hostlens}. ``S'' stands for the best-fit magnitude of the de-lensed QSO host galaxy, for which a de Vaucouleurs profile is used. Visits: (1) UT 2017 February 26, (2): UT 2017 May 7. Here, ``w/ GX'' stands for the lens mass model which accounts for GX as a perturber, whereas ``w/o GX'' stands for the mass model without a perturber. See Section \ref{sec:lensing} for details.}
\label{tab:photometry} 

\end{deluxetable*}

\begin{deluxetable*}{ccccc}



\tablewidth{0pt}

\tablecaption{Relative astrometry of W2M~J1042+1641}

\tablenum{4}

\tablehead{
\colhead{} & \multicolumn{2}{c}{Model w/ GX} &  \multicolumn{2}{c}{Model w/o GX} \\
\colhead{Component} & \colhead{E $\to$ W} & \colhead{S $\to$ N} & \colhead{E $\to$ W} & \colhead{S $\to$ N} \\ 
\colhead{} & \colhead{($\arcsec$)} & \colhead{($\arcsec$)} & \colhead{($\arcsec$)} & \colhead{($\arcsec$)} 
} 

\startdata
 A  & \phs$0.000\pm0.0004$ &  \phs$0.000\pm0.0004$ & \phs0.000$\pm$0.0004 & \phs0.000$\pm$0.0004 \\
 B  & \phs$0.151\pm0.005$ &       $-0.561\pm0.008$ & \phs0.147$\pm$0.006 & $-0.566\pm0.006$ \\
 C  & \phs$0.812\pm0.005$ &       $-0.911\pm0.006$ & \phs0.812$\pm$0.004 & $-0.913\pm0.006$ \\
 D  & \phs$1.590\pm0.005$ & \phs$0.539\pm0.009$ & \phs1.593$\pm$0.005 & \phs0.536$\pm$0.009 \\
 G  & \phs$0.777\pm0.004$ &       $-0.079\pm0.003$ & \phs0.775$\pm$0.002 & $-0.077\pm0.003$ \\
G1 &       $-2.140\pm0.009$ &      $-2.642\pm0.004$ &       $-2.140\pm0.010$ & $-2.643\pm0.005$ \\
GX &       $-0.841\pm0.029$ &      $-0.403\pm0.031$ &       $-0.827\pm0.024$ & $-0.399\pm0.021$ \\ 
\enddata


\tablecomments{Similar to Table \ref{tab:photometry}, ``w/ GX'' stands for the lens mass model which accounts for GX as a perturber, whereas ``w/o GX'' stands for the mass model without a perturber. We report the medians and the standard deviations of the values measured in the two filters, in both visits, relative to image A. For image A itself, we report representative MCMC uncertainties.}
\label{tab:astrometry} 

\end{deluxetable*}


\begin{deluxetable}{cccc}



\tablewidth{0pt}

\tablecaption{Lensing mass models}

\tablenum{5}

\tablehead{
 & \colhead{free SIE+$\gamma$} & \colhead{SIE+$\gamma$} & \colhead{SIE+$\gamma$+GX}
} 
\startdata
$\sigma_G$ & $222.0^{+0.9}_{-0.6}$ & 229.0$\pm$0.8 & $223.2^{+1.3}_{-1.1}$ \\
$e=1-b/a$ & $0.17^{+0.14}_{-0.08}$ & 0.55$\pm$0.02 & $0.43^{+0.04}_{-0.03}$ \\ 
$\theta_e$ & $49.2^{+8.7}_{-3.8}$ & 28.0$\pm$1.0 & $24.9^{+1.8}_{-2.2}$ \\ 
$\gamma$ & 0.03$\pm$0.01 & 0.15$\pm$0.01 & $0.08^{+0.02}_{-0.01}$ \\ 
$\theta_\gamma$ & $-5.7^{+25.0}_{-23.7}$ & $-66.7\pm1.5$ & $-75.5^{+5.3}_{-4.3}$  \\ 
$\sigma_{GX}$ & $-$ & $-$ & $46.5^{+4.2}_{-3.9}$ \\ 
$\Delta t$ (days) &  $12.9^{+7.4}_{-3.9}$ & 26.1$\pm$1.1 & 22.7$\pm$1.5 \\ 
$\chi^2/\nu$ & $0/-1$ & 62.5/1 & 0/0 \\ 
\enddata


\tablecomments{Based on fitting with \texttt{glafic}. Uncertainties are determined from 5 MCMC runs with $10^5$-$10^6$ steps each, for each mass model. We assume astrometric errors of $0.001\arcsec$ for image A, and for the other images we use the errors reported in Table \ref{tab:astrometry}.\footnote{For the SIE+$\gamma$ model we use uncertainties 10 times smaller than measured. Otherwise, we find that in order to match the observed position of the lensing galaxy, the predicted QSO images deviate too much from the observed positions. As a consequence, although the reported $\chi^2$ for this model was computed using the observed error bars, the error bars on the parameters may be artificially small.} Both G and GX are assumed to be at $z_l=0.599$ (see \S \ref{sec:lensz}), and the source is fixed at $z_s=2.517$. Angle are positive E of N. $\nu$ is the number of degrees of freedom. Image D leads, and all other images have similar time delays with respect to it, with differences of $\lesssim$ 1 day.}
\label{tab:lens} 

\end{deluxetable}


\begin{deluxetable}{cccccccc}



\tablewidth{0pt}

\tablecaption{Convergence and shear at the location of each QSO image}

\tablenum{6}

\tablehead{& \multicolumn{2}{c}{free SIE+$\gamma$} & \multicolumn{2}{c}{SIE+$\gamma$} & \multicolumn{2}{c}{SIE+$\gamma$+GX} & \\
\colhead{Image} & \colhead{$\kappa$} & \colhead{$\gamma$}  & \colhead{$\kappa$} & \colhead{$\gamma$}  & \colhead{$\kappa$} & \colhead{$\gamma$} & \colhead{$\kappa_\star$} 
} 
\startdata
A & 0.536 & 0.514 & 0.465 & 0.563 & 0.496 & 0.551 & 0.053 \\
B & 0.490 & 0.480 & 0.400 & 0.543 & 0.454 & 0.493 & 0.045 \\
C & 0.517 & 0.540 & 0.641 & 0.534 & 0.639 & 0.545 & 0.078 \\
D & 0.425 & 0.418 & 0.310 & 0.455 & 0.343 & 0.402 & 0.028 \\
\enddata


\tablecomments{To compute $\kappa_\star$ we first modeled the system with a De Vaucouleurs \citep{devaucouleurs48}+$\gamma$+GX mass profile using as priors the morphological parameters from Table \ref{tab:morphology}. We then reduced the mass in the best-fit De Vaucouleurs profile to a stellar fraction of 0.2 \citep{dai10}, and computed the resulting convergence at the location of the QSO images.}
\label{tab:kappagamma} 

\end{deluxetable}


\begin{deluxetable*}{lccccccccc}



\tablewidth{0pt}

\tablecaption{Morphology of the lens and nearby galaxy}

\tablenum{7}

\tablehead{
\colhead{Object \& Filter} & \colhead{$n$} & \colhead{$R_e$ (\arcsec)} & \colhead{b/a} & \colhead{PA (deg)}
} 

\startdata
G F125W (w/ GX) & 4.38$\pm$0.02 & 1.08$\pm$0.01 & 0.66$\pm$0.00 &  27.40$\pm$0.12 \\ 
G F125W (w/o GX) & 4.70$\pm$0.03 & 1.17$\pm$0.01 & 0.67$\pm$0.00 &  26.24$\pm$0.16 \\ 
G F160W (w/ GX) & 4.42$\pm$0.02 & 1.00$\pm$0.01 & 0.66$\pm$0.00 &  [27.40$\pm$0.12] \\ 
G F160W (w/o GX) & 4.74$\pm$0.02 & 1.03$\pm$0.01 & 0.66$\pm$0.00 &  [26.24$\pm$0.16] \\ 
G1 F125W (w/ GX) & 1.58$\pm$0.09 & 0.22$\pm$0.00 & 0.51$\pm$0.02 &  67.56$\pm$0.60 \\ 
G1 F125W (w/o GX) & 1.56$\pm$0.10 & 0.23$\pm$0.00 & 0.50$\pm$0.01 &  68.05$\pm$0.65 \\
G1 F160W (w/ GX) & 1.57$\pm$0.08 & 0.24$\pm$0.01 & 0.43$\pm$0.01 &  [67.56$\pm$0.60] \\ 
G1 F160W (w/o GX) & 1.53$\pm$0.07 & 0.25$\pm$0.01 & 0.41$\pm$0.01 &  [68.05$\pm$0.65] \\ 
\enddata


\tablecomments{Morphology has been measured with \texttt{hostlens}. Similar to Tables \ref{tab:photometry} and \ref{tab:astrometry} ``w/ GX'' stands for the lens mass model which accounts for GX as a perturber, whereas ``w/o GX'' stands for the mass model without a perturber. Angles are positive E of N. The effective radius is measured along the semimajor axis. For each filter, the modeling was done by enforcing the match of each morphological parameter between the two observing epochs. The position angle was also enforced to match between the two filters (represented here by the use of square brackets for F160W).}
\label{tab:morphology} 

\end{deluxetable*}


\subsection{Lensing analysis}\label{sec:lensing}

The relative astrometry of the four QSO images and of the lensing galaxy, from the HST data, provide the most robust constraints to determining a gravitational lens model for W2M~J1042+1641.  The procedure outlined in \S \ref{sec:coupledmodeling} is applied to two different lensing models, which we will explain later in this section. However, we will first analyze the ``definitive'' lens models constructed using the relative astrometry obtained by the iterative modeling described in the previous section and listed in Table \ref{tab:astrometry}. This will serve to motivate the choice of the two lens models mentioned above.

A commonly used model to fit gravitationally lensed QSOs, when the main constraints are astrometric, is the singular isothermal ellipsoid with external shear (SIE$+\gamma$). In this model the `strength' of the lens is characterized by a velocity dispersion, expected to be close to the central velocity dispersion of the stars in the lensing galaxy \citep[e.g.,][]{Kochanek94}\footnote{The other parameters of the model are: 1-2) the coordinates of the lensing galaxy, 3-4) the coordinates of the source, 5-6) the orientation of the major axis of the ellipsoid (projected on the plane of the sky), as well as its axis ratio, and 7-8) the orientation and strength of the external shear.}. This is one of two types of lens models we used in \S \ref{sec:coupledmodeling} to fit the imaging data, and we report its best-fit parameters in Table \ref{tab:lens}. However, this model results in a statistically very poor fit with $\chi^2\sim62.5$ for a single degree of freedom. The reason for the poor fit is that the model is unable to reproduce the observed locations of the QSO images. If we remove the constraint on the lens location, we obtain a perfect fit, although the model becomes under-constrained. We refer to this model as ``free SIE$+\gamma$'' in Table \ref{tab:lens}. 

Inspired by the work on lens galaxy environments by \citet{Sluse12b}, we next looked at the nearby environment of the system for clues that might explain the poor fit of our SIE$+\gamma$ model. Figures \ref{fig:hst1} and  \ref{fig:hst1_resid} reveal two structures near the lensing galaxy G: galaxy G1, located $3.90\arcsec$ from G, and GX, a structure much fainter but closer to the system ($1.67\arcsec$ from G and $0.94\arcsec$ from A). Including G1 in the fit as a second singular isothermal sphere (SIS) does not result in a significant improvement. In fact, its impact on the model based on its luminosity compared to that of the elliptical lensing galaxy and scaled by the Faber-Jackson law \citep{faber76} is negligible, and expected to be even smaller in reality, since it has a S\'{e}rsic index of $n \simeq 1.5$ suggestive of spiral morphology (see Table  \ref{tab:morphology}).
On the other hand, GX is a compact object whose morphology we are unable to resolve, but whose existence as a real object as opposed to a PSF artifact is validated by its presence at the same location in both filters and both visits. 
If we include it in the fit as a SIS at the observed location and at the redshift of the lens, with a velocity dispersion free to vary, we obtain a perfect fit for zero degrees of freedom (see Table \ref{tab:lens}). This is the second type of model we used for the iterative fitting in \S \ref{sec:coupledmodeling}.

We note that, in a program to study a sample of 30 lensed QSOs with HST (Cycle 26, Program ID 15652, PI: Treu), W2M J1042+1641 was imaged in the F475X and F814W filters with WFC3/UVIS. \citet{Schmidt22} modeled this system as part of the sample using an automated pipeline using the {\scshape Lenstronomy} software package \citep{Birrer18} with a power law elliptical mass distribution plus external shear. The fitting was done simultaneously for the two UVIS bands plus the F160W data from this work, and the F125W data were not used. The astrometric positions derived for the QSO components differ from what we find here by up to $\sim40$ mas, which is still at the sub-pixel level, but results in a different lens model. The automated nature of this analysis does not include the finer structures such as sources GX and X and, in addition, the reported astrometry is fine-tuned to the specific choice of lensing model (private communication), making a direct comparison to our results infeasible.

\subsection{The Fiducial Model} 

\begin{figure}[ht!]
\epsscale{1.4}
\plotone{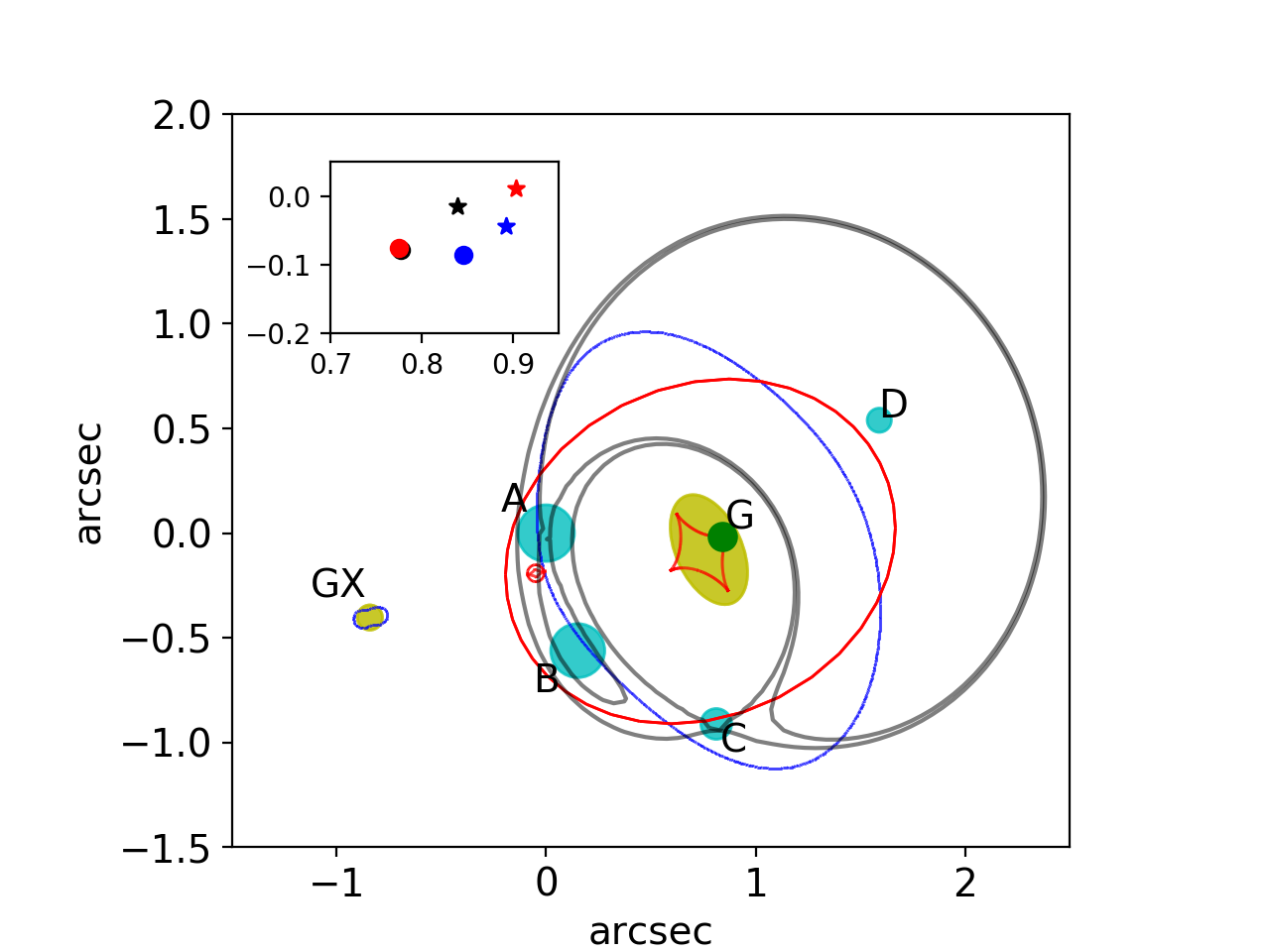}
\caption{Lensing configuration for the SIE$+\gamma$+GX model. The two lenses (G and GX) are marked with yellow ellipses whose sizes scale in proportion to the modeled velocity dispersions. The ellipsis for G shows the orientation and axis ratio predicted by the mass model. Critical lines are drawn in blue and caustics in red. The green circle marks the location of the source QSO in the source plane, and the cyan circles mark the locations of the observed images. The size of the circles is in proportion to the predicted flux ratios. 
{\bf Time-delay surface isochrones at the location of the saddle point images A and C are drawn in gray.}
In the inset, the source QSO position (stars) and lens position (circles) are marked for the SIE$+\gamma$+GX model (black), the SIE$+\gamma$ model (red), and the free SIE$+\gamma$ model (blue). The black circle coincides with the light profile centroid of the lensing galaxy.
}\label{fig:lens} 
\end{figure}

In addition to the quality of the fit, we list the following arguments as to why the SIE$+\gamma$+GX model is more realistic: 

\begin{enumerate}
\item While the velocity dispersion of GX was a free parameter during the fit, its best-fit value of $\sim 46.5$ km/s is in good agreement with the predicted value of $\sim50$ km/s based on its relative luminosity compared to G, and the Faber-Jackson law. Though we can only estimate a rough photometric redshift for GX based on a single color, it is in good agreement with our best estimate for G (\S \ref{sec:lensz}), but only if we assume a late-type galaxy spectral template (GX appears bluer than G in Figure {\bf \ref{fig:hst1_resid}}).

\item Previous lensing studies find that there is a good alignment between the axes of the light and mass distributions in lensing galaxies, within $\sim 10$ deg \citep[e.g.,][]{Shajib19,Keeton98,Sluse12b}. We find that when GX is added to the model, the mass and light profiles of the main lens G do indeed show excellent alignment (the light PA for G in Table \ref{tab:morphology} is $\sim 27\degr$ and the mass profile $\theta_e = 24.9\degr$ in Table \ref{tab:lens}), while the models without GX show misalignment by $\sim20\deg$ ($\theta_e = 45\degr-49\degr$ in Table \ref{tab:lens}). This supports the SIE$+\gamma$ and the SIE$+\gamma$+GX models over the free SIE$+\gamma$ model.

\item The best-fit location of the lensing galaxy in the free SIE$+\gamma$ model is $0.072\arcsec$ away from the center of the observed light profile, towards west (see inset in Figure \ref{fig:lens}). However, previous work by \citet{Yoo2006} found that typical offsets between the mass and light centroids is on the order of a few mas. This result was further confirmed by \citet{Sluse12b}, which finds similar values for 11/14 quads, once they include nearby luminous perturbers in their models.\footnote{We note that \citet{Shajib19} find offsets larger by about a magnitude in their modeling of 13 quads with {\em HST} imaging, however their exploratory models do not account for potential nearby perturbers.} As with the previous point, this supports the SIE$+\gamma$ and the SIE$+\gamma$+GX models over the free SIE$+\gamma$ model. It should be noted, however, that not only is the SIE$+\gamma$ a poor fit, but its parameters are a more extreme version of those of the SIE$+\gamma$+GX model. In particular, from Table \ref{tab:lens}, both the ellipticity of the lens ($e$) and the value of the shear ($\gamma$) have increased, and their directions are within 5 deg of being perpendicular, a potential sign of degeneracies in the model.

\item Like any cusp configuration in a smooth lensing mass model, the central image (in this case B) is expected to have a flux equal to the sum of the two surrounding images \citep[A and C; e.g.,][]{Keeton03}. Depending on which filter/visit is used to measure the observed flux, this makes the observed flux of image A, in particular, anomalous by more than an order of magnitude compared to the SIE$+\gamma$ model (see \ref{sec:magnif}). Such an anomalous flux ratio exceeds the largest previously recorded anomaly in the optical, for SDSS~J0924+0219 \citep{Inada03}.\footnote{In the case of SDSS~J0924+0219, the anomalous flux is suppressed rather than enhanced, in contrast to the present case.} The addition of GX boosts the predicted flux of image A in relation to B by a factor of $\sim2$ compared to the free SIE$+\gamma$ model, and by a smaller amount compared to the SIE$+\gamma$ model, thus mitigating some of the discrepancy (see \S \ref{sec:magnif} for a comprehensive analysis of the flux ratios).
\end{enumerate}

We therefore adopt the SIE$+\gamma$+GX model as our fiducial model, which we used to fit the morphology in Figures \ref{fig:hst1} and \ref{fig:hst1_resid}, and for which we plot the image configuration, critical lines, caustics and time delay surface in Figure \ref{fig:lens}. We show the equivalent of Figures \ref{fig:hst1} and \ref{fig:hst1_resid}, but employing the free SIE$+\gamma$ model, in Appendix \ref{SIEgamma}. We note that we do not find the difference in the residuals after fitting these two models with \texttt{hostlens} to be large enough to rule out one of the models, so our choice of the fiducial model is based entirely on the four arguments given above. 

\section{Discussion}\label{sec:discussion}

With our fiducial (SIE$+\gamma$+GX) model in hand, we explore in this section the unique properties of this system.  We investigate possible explanations for the flux anomaly seen among the QSO components. We also re-visit the nature of this object as a red QSO and study its black hole and host galaxy properties in the context of their co-evolution.

\subsection{Flux ratio analysis and total magnification}\label{sec:magnif}

\begin{figure*}
\epsscale{1.25}
\plotone{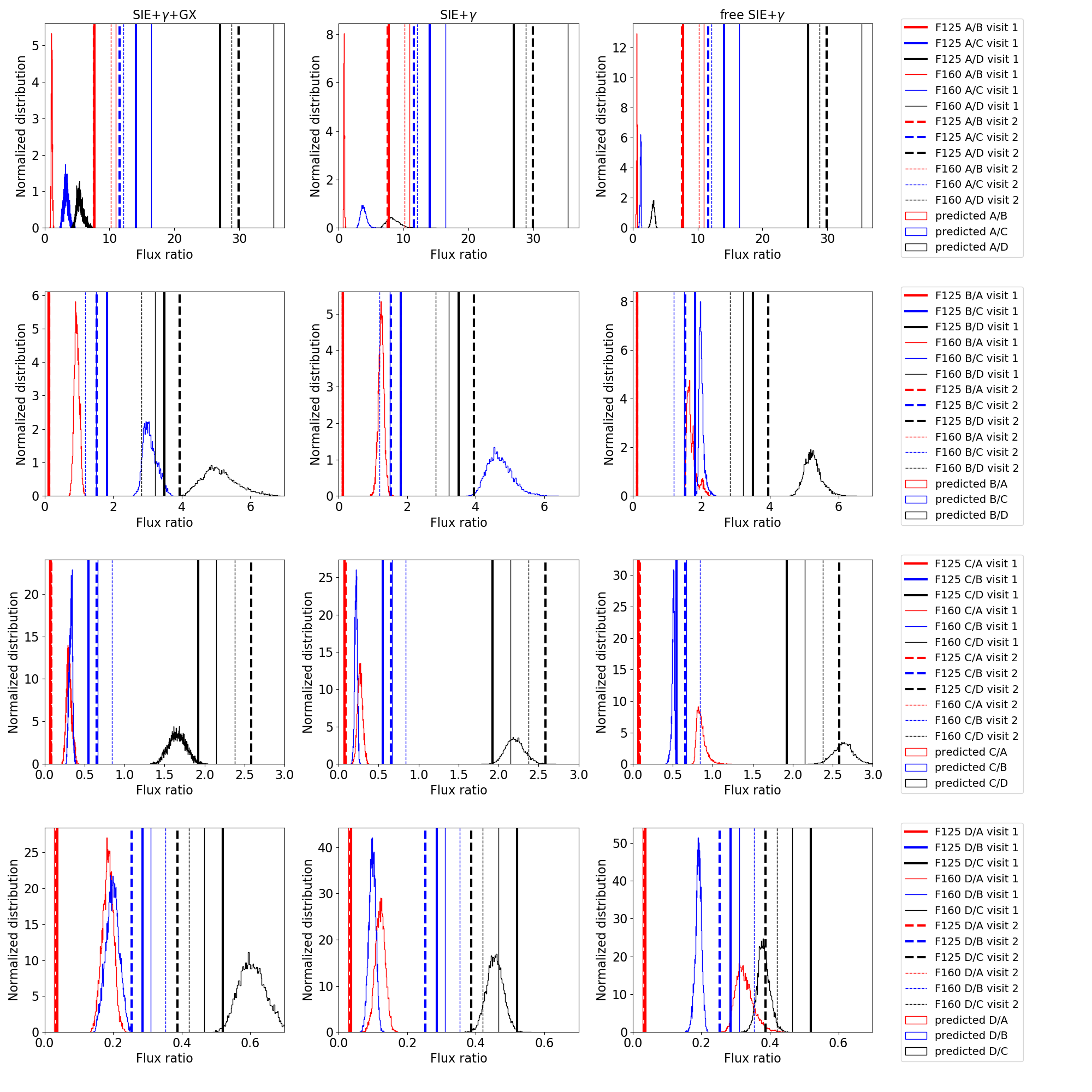}
\caption{Comparison of observed and predicted flux ratios. The lens mass models, from left to right, are SIE$+\gamma$+GX, SIE$+\gamma$ and free SIE$+\gamma$. The histograms consist of 10000 points from the MCMC chains computed with \texttt{glafic}, and the vertical lines correspond to the photometry measured by \texttt{hostlens}. Visits: (1) UT 2017 February 26, (2): UT 2017 May 7. The fluxes of image D in visit 2 may be unreliable due to a diffraction spike from image A falling on top of it.
}
\label{fig:fluxratio} 
\end{figure*}

A crucial quantity needed for the purpose of characterizing the physical properties of the source QSO is its total magnification. 
However, due to the large flux anomalies present in this system, this quantity is not trivial to determine. 
We show these anomalies in Figure \ref{fig:fluxratio}, by comparing the measured flux ratios with the histograms of the ratios of image magnifications, corresponding to the lens models from Table \ref{tab:lens}.\footnote{The magnification histograms for each image, which go into the flux ratios in this figure, as well as into the subsequent discussion on the total magnification, were computed using the MCMC chains used in Table \ref{tab:lens}, which assumed a point source. However, we can check that they hold for a realistic size of the accretion disk as follows. The standard deviation of the source position with respect to the lens position, obtained from MCMC, is $\sim12$ mas (for the SIE$+\gamma$+GX mas model). At the source redshift, this corresponds to $\sim100$ pc. However, the accretion disk size we estimate in \S\ref{sec:microlens} is $\sim100$ light days, in agreement with estimates of the size of the broad-line region in the literature \citep[e.g.,][]{gravity18}. Since this is on the order of a thousand times smaller than the measured standard deviation of the source position, we conclude that the physical size of the source has no effect on the magnifications and flux ratios.} The fiducial lens model, SIE$+\gamma$+GX, is shown in the left column of Figure \ref{fig:fluxratio}, the middle column shows the SIE$+\gamma$ model, and the right-hand column shows the free SIE$+\gamma$ model. 
As we noted in \S \ref{sec:lensing}, all flux ratios related to image A are highly anomalous.
To compute a robust magnification, we require at least one match between an observed and a model-predicted flux ratio. 

\begin{figure}
\epsscale{1.1}
\plotone{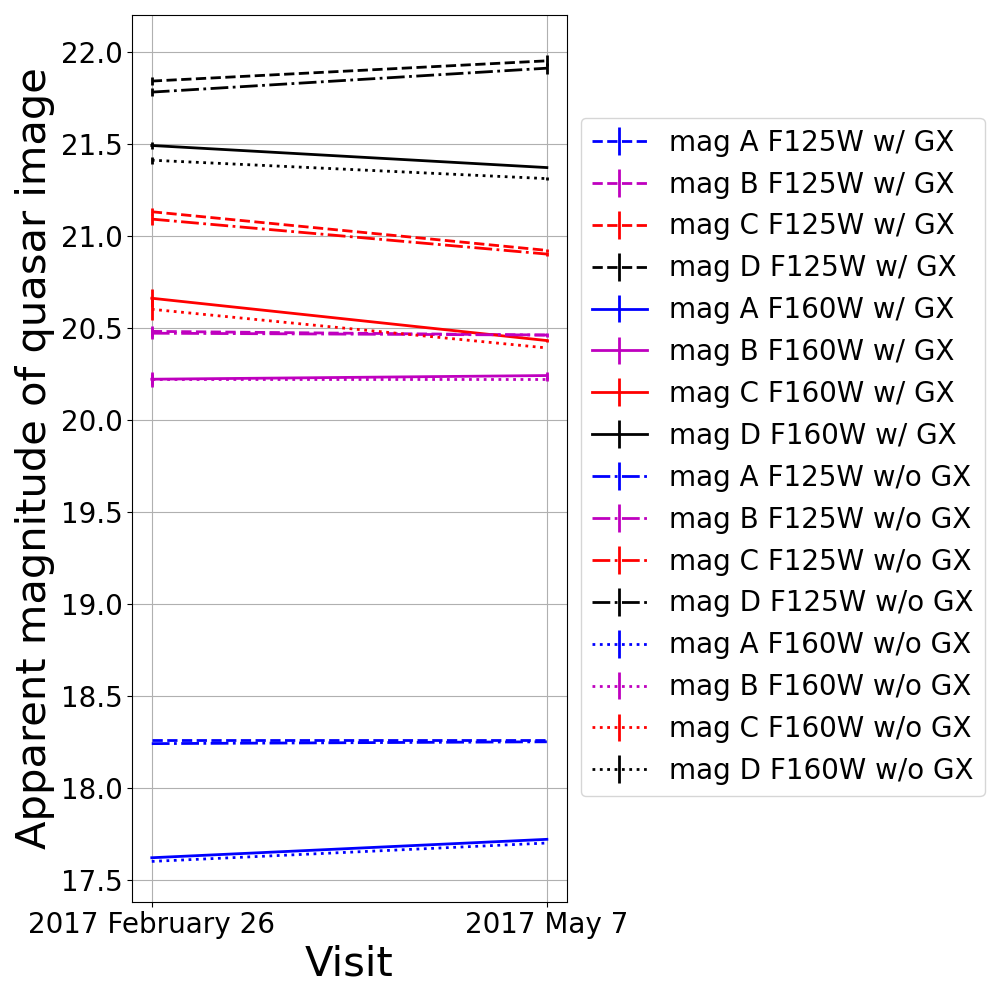}
\caption{Variation of the QSO image magnitudes with observing epoch.
}\label{fig:magchange} 
\end{figure}

The least anomalous flux ratios are those involving images C and D. The observation-prediction overlap in Figure \ref{fig:fluxratio} is small for the fiducial model, but larger for the free SIE$+\gamma$ model and even larger for the SIE$+\gamma$ model. We note, however, the following caveats of the observed flux ratios: First, figure \ref{fig:hst1} shows that in both filters of visit 2, image D is covered by a diffraction spike of image A. It is difficult to assess what level of systematics this may introduce for the photometry of image D, reported in Table \ref{tab:photometry}. 
A clue that the photometry of D might be problematic is that Figure \ref{fig:magchange} shows different directions of variation for the magnitude of image D in the two filters, as opposed to image C which varies in a single direction, and also against expectations if the change was caused by microlensing or intrinsic variability. 
By comparing with the direction of variation of image C, the flux of D in F125W visit 2 is more likely to be affected by systematics.
Second, both microlensing and intrinsic variability would cause variations of smaller amplitude at longer wavelengths, so the flux ratios in F160W should give a better estimate of the intrinsic flux ratio. 

Based on the arguments we presented in \S \ref{sec:morphmodeling}, we construct an ``anchor'' by using the fiducial lens model, in spite of the fact that the SIE$+\gamma$ model shows a better match, and intersecting the C/D model prediction distribution with a Gaussian fit to the observed fluxes in filter F160W only\footnote{We have a direct measurement of C/D $=1.05\pm0.34$ from ALMA imaging of host galaxy thermal dust emission in \S\ref{sec:alma}, which is lower by a factor of $\sim2$ than the {\it HST}-derived value, as well as the lens model predictions. However, this sub-mm emission is extended (see Figure \ref{fig:radio}), and expected to be primarily associated with dust in the QSO host galaxy \citep[see \S\ref{sec:fluxradio};][]{stacey18,stacey21}. These characteristics make the ALMA-based measurement unsuitable to use as an anchor.}. We checked that, although this intersection samples from the tail of the C/D model prediction distribution, the resulting samples correspond overall to the distribution of $\chi^2$ values of the entire MCMC chain we used in Table \ref{tab:lens}, and therefore not to particularly poor fits. Finally, the total magnification we compute, i.e., integrated over the flux of the four images, accounting for flux ratio anomalies, is\footnote{If we include the observed flux ratios in F125W when computing the intersection we obtain $107^{+15}_{-15}$. If we compute $obs_\mathrm{A/D}\times|\mu_{\mathrm{D},C/D\cap}|+obs_\mathrm{B/D}\times|\mu_{\mathrm{D},C/D\cap}|+|\mu_{\mathrm{C},C/D\cap}|+\mu_{\mathrm{D},C/D\cap}$ instead, we obtain $145^{+16}_{-20}$. Alternatively, we get $obs_\mathrm{A/B}\times\mu_{\mathrm{B},B/C\cap}+\mu_{\mathrm{B},B/C\cap}+\mu_{\mathrm{C},B/C\cap}+obs_\mathrm{D/B}\times\mu_{\mathrm{B},B/C\cap}=221^{+23}_{-21}$. Finally, alternatively, if we use the lens model free SIE$+\gamma$, we obtain $196^{+90}_{-57}$.}

\begin{multline}
\mu_\mathrm{total}^\mathrm{anomalous A, B}=obs_\mathrm{A/C}\times|\mu_{\mathrm{C},C/D\cap}|+obs_\mathrm{B/C}\times|\mu_{\mathrm{C},C/D\cap}| \\
+|\mu_{\mathrm{C},C/D\cap}|+\mu_{\mathrm{D},C/D\cap}=117^{+22}_{-14}.
\label{eq:mu} 
\end{multline}

\noindent Here $obs$ refers to the observed flux ratios and $\mu$ is the model-predicted magnification for each individual image. $C/D\cap$ denotes that the magnifications are subject to the intersection condition introduced above.  

The total magnification computed above is what we would use to study the physical properties of this system based on on its observed, unresolved, longer-wavelength data, if we expect that the flux anomalies we see in the {\em HST} images still hold at those wavelength, and the moment in time when those data were collected. On the other hand, if images A and B were not anomalous, the fiducial model predicts a total magnification of\footnote{Removing the intersection constraint, the total magnification is $49.0^{+3.7}_{-3.1}$.} \footnote{Following the argument in Appendix \ref{sec:host}, the uncertainty on the radial slope of the host also introduces an additional uncertainty of $\sigma_\mu/\mu\sim0.24$ on the total magnification, here as well as in the paragraph above.}

\begin{multline}
\mu_\mathrm{total}^\mathrm{predicted A, B}=(|\mu_{\mathrm{A}}|+\mu_{\mathrm{B}}+|\mu_{\mathrm{C}}|+\mu_{\mathrm{D}})_{C/D\cap}=52^{+5}_{-4}.
\end{multline}

\noindent This is what we must use if we expect the flux anomalies to disappear at longer wavelengths. 
In the following sections, we explore one by one the three physical phenomena known to be responsible for flux anomalies in general.

\subsubsection{Extinction}\label{sec:extinct}

In addition to the intrinsic reddening that the QSO experiences from dust in its own host galaxy, discussed in \S \ref{sec:spec1}, we consider the possibility that the different lensed lines of sight may be reddened as well.
To date, the largest sample of lensed QSOs (23 systems) used to study the extinction properties in these systems remains the one of \citet{Falco99}. The median differential extinction was found to be $E(B-V)=0.04$, the median total extinction $E(B-V)=0.08$, and the median $R_V$ in particular, for the early-type sample, was consistent with the Galactic value of 3.1. The consistency with the Galactic extinction parameter was later confirmed by \citet{Eliasdottir06,Ostman08}). Using the extinction curve from \citet{Cardelli89}\footnote{We perform the calculation using \texttt{extinction} (\url{https://extinction.readthedocs.io})}, for the assumed lens redshift in \S \ref{sec:lensz}, we find median differential extinction of 0.08 mag in F125W, 0.05 mag in F160W and median total extinction of  0.15 mag in F125W, 0.10 mag in F160W. The differential extinction is too small to significantly impact our measured flux ratios, and accounting for total extinction would only correct the inferred total magnification upward by 10-15\%. Because these median values are small and their parent distributions are broad, we do not implement extinction corrections. 

We note that the lens galaxy in W2M~J1042+1641 is early-type, thus expected to be dust- and gas-poor, hence producing smaller extinction. While \citet{Falco99} do not find a correlation with the impact parameter, we report that the effective radius is $\sim1\arcsec$, and the QSO images are located at an impact parameter between $0.78\arcsec - 1.02\arcsec$, with images C and D located farthest away. Finally, we note that \citet{Peng06} also found that extinction is small enough to ignore in their study of the black hole - QSO host coevolution from a sample of lensed QSOs. 

\subsubsection{Microlensing analysis, intrinsic variability and time delays}\label{sec:microlens}

\begin{figure*}
\epsscale{1.3}
\plotone{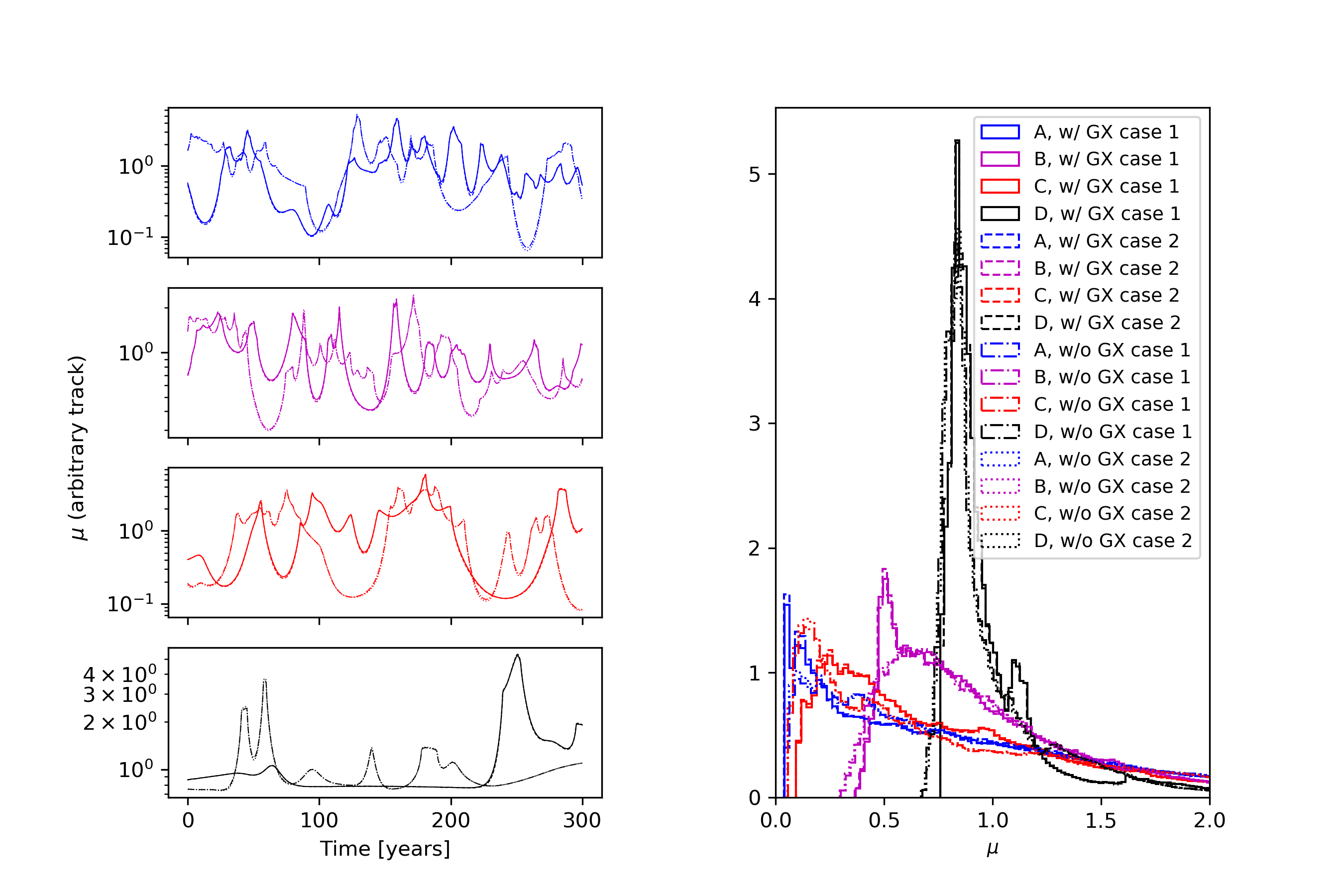}
\caption{Simulated microlensing light curves in F160W. {\it Left}: Magnification curves along an arbitrary track of the magnification maps of the four QSO images. The magnification of each image in the absence of microlensing has been normalized to unity by dividing over the magnifications predicted by two best-fit lensing models, SIE$+\gamma$+GX and SIE$+\gamma$ with free lens position. The time scale is chosen to show the typical frequency of microlensing events for each image. {\it Right}: Histograms of the microlensing magnification over the entire microlensing map. The histograms continue to decrease smoothly beyond the limit displayed. {\it Case 1}: $\log M_{BH}[M_\sun]=9.06$, $L/L_{Edd}=0.21$, $\eta=0.43$, $R_0=$2.76e+15 m; {\it Case 2}: $\log M_{BH}[M_\sun]=9.56$, $L/L_{Edd}=0.046$, $\eta=1.23$, $R_0=$2.52e+15 m. Both cases use 1.0$R_0$, $i=60\deg$, $PA=90\deg$. See \S\ref{microlensdetails} for further details.
}\label{fig:microlensing} 
\end{figure*}

We conducted microlensing simulations in order to investigate the plausibility of microlensing as the reason for the large flux anomaly of image A. We report the relevant values of convergence and shear at the location of each image in Table \ref{tab:kappagamma}, for all three best-fit lens mass models. We show the details of the computation, for which we employ the black hole properties derived in \S\ref{sec:mbh}, in Appendix \ref{microlensdetails}. We show our results for the SIE$+\gamma$+GX and SIE$+\gamma$ mass models in Figure \ref{fig:microlensing}. In the left-side panel, we show for each of the four images the microlensing magnification for a random microlens track relative to the source image, where the base value of zero corresponds to the magnification value of the fiducial macro-lens mass model. We plot two of our simulated cases, corresponding to two different sets of physical properties ($M_{BH}$, $L/L_{\rm Edd}$) which we derive in \S\ref{sec:mbh}. In the right-side panel we show the histograms of the magnifications over all simulated tracks. From these histograms we find that images A and C have the largest probability of being affected by microlensing, as they correspond to the broadest histograms, though most of the time they would be demagnified, being saddle point of the time delay surface, in agreement with, e.g., \citet[][]{Schechter02}. On the other hand, the flux of image D is the least impacted by microlensing. However, the probability that image A is magnified by a factor as large as required to explain the observed anomaly is only $0.22^{+0.55}_{-0.18}\%$ (for the SIE$+\gamma$+GX model, in case 2 from the plot; $0.17^{+0.48}_{-0.14}\%$ for case 1. For the SIE$+\gamma$ model the probability is even smaller). We therefore conclude that, while microlensing can explain part of the observed anomaly, it is highly unlikely to explain all of it. Flux changes this large would take on the order of a decade to complete. Finally, we note that while the integrated spectrum does show an enhancement in the flux in the blue, we have argued in \S \ref{sec:spec1} that this can be explained by factors intrinsic to the QSO, and does not require chromatic microlensing.

Since the QSO images correspond to the same physical source, if we ignore microlensing, any variation due to intrinsic variability has to be reflected in all images, shifted in time by the time delays. The order of arrival of the images is D, B, A and C. The time delay between images A, B and C is $\lesssim1$ day, while between D and the other images it is $\sim23$ days. As the time delay between images A, B and C is so short, we do not expect to see variations due to intrinsic variability between these images, whose amplitude are expected to be governed by the structure function \citep[e.g.,][]{vandenberk04}. Indeed, the flux ratio A/B is preserved at two epochs, as we showed in \S \ref{sec:radio}. Nevertheless, Figure \ref{fig:magchange} shows variations in images A (blue lines; $\sim0.1$ mag) and C (red lines; $\sim0.2$ mag) on the timescale of 70 days (20 days in the QSO rest frame). Because the flux of image B (magenta lines) is constant during this timescale, we interpret this to imply the lack of QSO intrinsic variability (or at least to imply variations resulting in the same start and end point), and also as a check of our absolute photometric calibration (which we have also checked independently, finding consistent photometry for the bulk of the sufficiently bright objects in the {\it HST} field of view). More puzzlingly, image A varies only in the longer wavelength filter, F160W, and image D (black lines) varies in opposite directions between the two filters, by $\sim0.1$ mag, with the variation in F160W tracking that of image C in both filters. We find these variations difficult to reconcile with microlensing, which affects shorter wavelengths more prominently, although microlensing can be responsible for variations of $\lesssim 0.15$ mag on a similar timescale \citep[see Figure 13 in][]{Eigenbrod08}. We note also that these variations are far outside the error bars we measure in Table \ref{tab:photometry} with our careful light profile fitting technique.  

In parallel to the microlensing simulations where we explored flux changes, we also looked into the effect that microlensing can have on QSO time delays, assuming the lamp-post model \citep[e.g.,][]{Tie18}. We find that the RMS impact on the time delays for images A, B, and C is as large as $\sim18$ days, $\sim7$ days and $\sim13$ days, respectively (for the largest accretion disk we considered, of size $2R_0$). These values are significantly larger than the $\lesssim1$ day intrinsic time delay between these images, and when plugged into the structure function (after conversion to rest-frame), they are large enough to factor in a hypothetical explanation of the ~0.1 mag stochastic variations mentioned above. It should be noted that while a larger disk size has an increased impact on the time delays, it also smoothes out the microlensing signal, decreasing the magnitude of flux ratio anomalies. We refrain from speculating on the exact mechanisms of these flux variations, as we do not have enough data to constrain them. 

\begin{figure}
\plotone{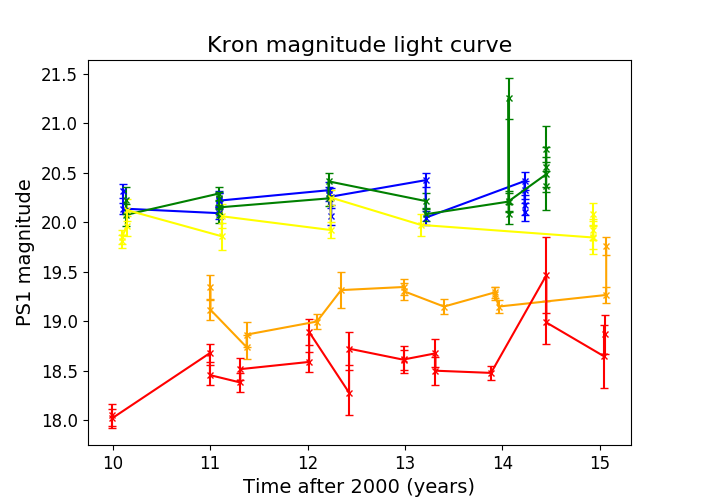}
\caption{Variability of W2M~J1042+1641 in individual Pan-STARRS1 exposures. Blue, green, yellow, orange and red colors represent, in order, $grizy$. Mismatches in photometry between points very close in time suggest that the photometric errors are underestimated, weakening evidence for variability. Note that the SDSS magnitudes from Table \ref{tab:photometrysurveys}, taken earlier in time, are in relatively good agreement. 
}\label{fig:ps1} 
\end{figure}

Finally, we have also looked for direct evidence of variability in the longer, 5-year baseline spanned by Pan-STARRS1 \citep{chambers16}, which we show for completeness in Figure \ref{fig:ps1}. However, the uncertainties of the automatic single-component fit to the photometry provided by the Pan-STARRS pipeline are likely to be underestimated, since the system consists of 4 QSO images and a bright lensing galaxy. We therefore can only conclude that there is no evidence for a monotonous variation recorded by Pan-STARRS1, that would correspond to long-term variability dominated by image A.

\subsubsection{Substructure in the lensing galaxy}\label{sec:substructlens}

\begin{figure}
\epsscale{1.3}
\hspace*{-0.5cm} 
\plotone{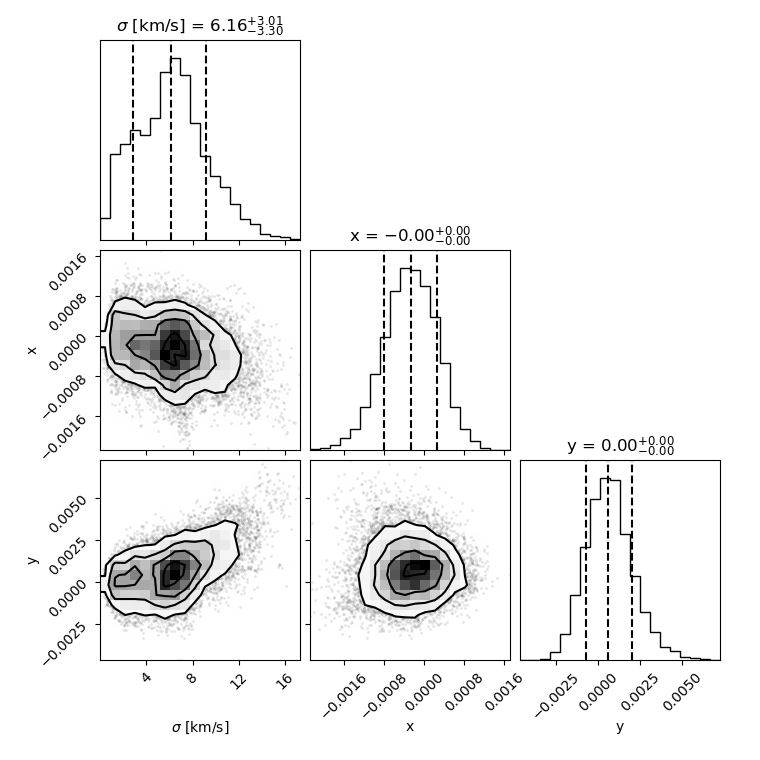}
\caption{Velocity dispersion and position (in arcsec) of a SIS substructure added to the SIE$+\gamma$+GX model around image A, in order to boost its flux by a factor of $\sim5$. The 16th and 84th velocity dispersion percentiles correspond to a mass (inside the substructure's Einstein radius) of $4.1\times10^3 M_\sun$ - $4.3\times10^5 M_\sun$.
}\label{fig:substruct} 
\end{figure}

$\Lambda$CDM substructure, either dark or luminous, has long been invoked to explain flux ratio anomalies in lensed quads \citep[e.g.,][]{dalal02}. If the flux anomalies we measure in the HST data are due to substructure, we expect them to persist at long wavelengths as well, as long as the emitting region is not very extended (i.e., AGN radio core emitting region, typically with a parsec-scale length, is small enough to be affected by substructure, but kpc-scale stellar emission is not). In Figure \ref{fig:substruct} we demonstrate with a toy model that a relatively small-mass perturber, placed on the order of milli-arcseconds away from image A, can produce enough magnification to cause the highly anomalous flux observed in this image, while leaving all other observables unchanged\footnote{We used 5 MCMC chains of 10000 points each. The fluxes of the images predicted by the SIE$+\gamma$+GX model were used as constraints (the flux of image A being boosted), with 5\% uncertainty. All other lensing parameters except those characterizing the substructure were held fix.}.
 
\subsubsection{Clues from the radio data}\label{sec:fluxradio}

Radio and far-IR photometry of lensed QSOs is considered to be unaffected by microlensing \citep[e.g.,][]{koopmans03}, as well as dust extinction. This is because the emission region in the source plane, on scales of milliarcsec, corresponds to radii much larger than the projected Einstein radius of individual stars in the lensing galaxy (scales of micro- or nanoarcsec). If the emission in radio and far-IR is due to the AGN, we would expect to measure a similarly anomalous flux ratio to what we measure with {\it HST} if it is the case that the anomaly is due to substructure (we use here the term to refer to a low-mass perturber), but we should measure something comparatively closer to the non-anomalous prediction of the best-fit lens model, if it is due to microlensing. We would not expect perfect matches to either of these values because 1) as mentioned above, the size of the emission region in the radio (as well as in the sub-mm) is not the same as that of the optical accretion disk; for example, the isophotes in Figure \ref{fig:radio} show extended structure. 2) the image separations measured from {\it HST} (Table \ref{tab:astrometry}) and from ALMA (Table \ref{tab:radio_mm}) are inconsistent, e.g., the separation between images A and B being $0.58\arcsec$ and $0.51\arcsec$, respectively, and the separation between the other images being even more discrepant. We note these caveats upfront, and point out that the data is not constraining enough to estimate to what extent they can affect the analysis we present below. 

The VLA images reveal a flux ratio of $A/B\sim4.0$, compared to the lens model prediction $A/B\sim1$ and the {\it HST} measurement of $A/B\sim6.6-10.2$ (depending on the measurement filter and visit). The integrated emission that we measure from the two epochs of VLA data is $\sim60-100\mu$Jy, $6-10$ times larger compared to the prediction of $\sim10\mu$Jy based on the ALMA data, assuming a typical effective dust temperature of 40 K \citep[see][]{stacey18,stacey19} and using the known far-IR $-$ radio correlation of star-forming galaxies \citep[e.g.,][]{kruit71,yun01,ivison10a,ivison10b}. The excess of measured flux at radio wavelengths suggests that we are seeing AGN emission in addition to synchrotron emission from star formation\footnote{Another source for enhanced radio emission in red QSOs is shocks from winds, as proposed in \citet{Glikman22}. }. 
This is in line with current models of QSO evolution, which suggest that coexistence of dust-obscured star formation and AGN activity is typical of most QSOs \citep[e.g.,][]{condon13,stacey18}. We can explain the mismatch in the {\it HST}-derived and radio/sub-mm -derived A/B flux ratios if the radio emission consists of a combination of point-like AGN emission and star formation-related emission. 
Furthermore, if the anomalous {\it HST} flux ratio is due to substructure; extended star formation emission is expected to be comparatively much less affected by a low-mass perturber, which preferentially magnifies the comparatively much more compact AGN emission. 

By accounting for the amount of AGN emission relative to the emission from star formation we can reproduce the intermediate flux ratio that we measure in the VLA data. Let $A_{\mathrm{SF}}$ and $B_{\mathrm{SF}}$ be the flux densities measured in the radio from star formation in images A and B, respectively, and let $A_{\mathrm{AGN}}$ and $B_{\mathrm{AGN}}$ be the flux densities measured from compact AGN emission. Using the constraints $A_{\mathrm{SF}}/B_{\mathrm{SF}}\sim1$, $A_{\mathrm{AGN}}/B_{\mathrm{AGN}}\sim6.6-10.2$, and $(A_{\mathrm{AGN}}+A_{\mathrm{SF}})/(B_{\mathrm{AGN}}+B_{\mathrm{SF}})\sim4.0$, we can solve for $A_{\mathrm{AGN}}/A_{\mathrm{SF}}\sim4.7-20.4$ and $B_{\mathrm{AGN}}/B_{\mathrm{SF}}\sim1.4-0.5$. Assuming that the difference between the ALMA-predicted flux densities for the VLA emission and the detected values is also due to further emission from AGN, we measure $(A_{\mathrm{AGN}}+A_{\mathrm{SF}}+B_{\mathrm{AGN}}+B_{\mathrm{SF}})/(A_{\mathrm{SF}}+B_{\mathrm{SF}})\sim6-10$, and we calculate a very consistent range of $\sim4-11.5$.

In light of the results above, we conclude that the radio data is more consistent with the {\it HST}-measured flux ratio anomaly of image A persisting at longer wavelengths, and therefore being mostly due to substructure rather than microlensing. Therefore, the total magnification to use for determining the source properties is the large value we computed in \S \ref{sec:magnif} ($\mu \sim 117$) from the observed flux ratios in the {\it HST} data and we adopt this value when de-magnifying fluxes to consider the QSO's intrinsic properties, in the sections that follow. 

Finally, we conclude this section with two notes. First, in the sub-mm emission measured by ALMA, A/B $=1.61\pm0.24$, which is somewhat larger than the model prediction of $\sim1$. One possible explanation would be that of thermal dust emission from a sufficiently compact region located close enough in projection to the substructure to be sufficiently magnified. Indeed, \citet{stacey21} showed that such emission can be as compact as a few hundred pc. Second, it is worth mentioning that similar analyses combining near-IR, sub-mm and radio flux-ratio measurements were recently performed by \citet{badole20} for the highly optically-anomalous SDSS J0924+0219, concluding that the long-persisting anomaly is most likely caused by microlensing; and by \citet{stacey18} for MG~J0414+0534, suggesting that the anomaly is caused by substructure. 

\subsection{Spectral characteristics}\label{sec:qso}

\begin{figure*}[ht!]
\plotone{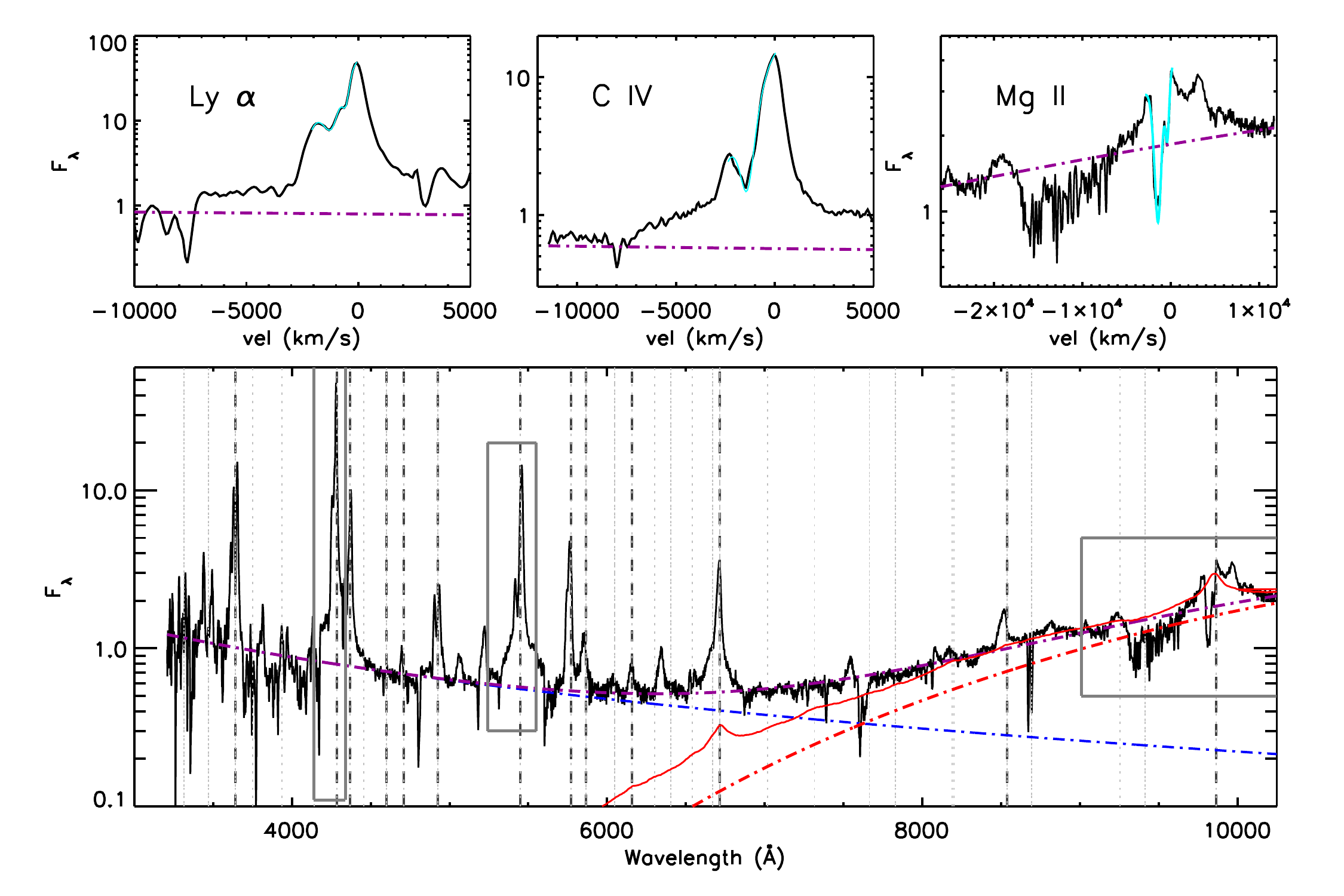}
\caption{
Deep LRIS spectrum of W2M~J1042+1641 (bottom panel; black line) with typical (light grey) and prominent (dark grey) QSO emission lines marked with vertical dotted and dashed lines, respectively.  A best-fit reddened power law continuum plus an unreddened leakage component is shown with a purple dash-dot line, with the reddened ($E(B-V)=0.89$) and leaked components plotted in red and blue, respectively.  The solid red line is the best-fit reddened template with $E(B-V)=0.73$.  
{\em Top row --} Profiles of prominent emission lines \ion{Mg}{2}, \ion{C}{4}, and Ly$\alpha$ (left to right) plotted in velocity space.  All three lines show a distinct absorption feature, that is well fit by a single or double Gaussian profile, with an outflow speed of $\sim 1000 - 1500$ km s$^{-1}$.
We note that the absorption feature blueward of \ion{Mg}{2} is due to telluric absorption.
The line regions plotted in the squares at the top are also marked with grey boxes in the bottom panel.
}\label{fig:LRIS} 
\end{figure*}

The LRIS spectrum of W2M~J1042+1641, taken across multiple PAs, for a total of 50 minutes (\S \ref{sec:lris}) is our best dataset for investigating the QSOs emission line properties.  Figure \ref{fig:LRIS} shows the combined LRIS spectrum of W2M~J1042+1641 (black line) plotted on a logarithmic y-axis in order to enhance features across its dynamic range.  

To identify and study line features, we require a better determination of the QSO continuum.  Following our interpretation that W2M~J1042+1641 is a reddened QSO with some leakage of the intrinsic spectrum, we model the QSO spectrum as a power law, with spectral index $\alpha_\nu = -0.5$ ($F_{\lambda0} \propto \lambda^{-(\alpha_\nu+2)}$). We then fit the line-free portions of the spectrum with a two-component power-law, one reddened and one pure, both with the same power-law index:
\begin{equation}
F_\lambda = A F_{\lambda 0}  + B F_{\lambda 0} e^{-\tau_\lambda}. \label{eqn:cont}
\end{equation}
The best fit is shown with a purple dash-dot line along with the unreddened component, plotted with a blue dash-dot line, and the reddened component, plotted with a red dash-dot line.  For comparison, we also show the best-fit reddened QSO template (red solid line) that was determined from the near-infrared spectrum combined with the LRIS spectrum, using only wavelengths longer than 8000\AA\ (\S \ref{sec:lris}).  While the template-based fit results in $E(B-V)=0.73$ mag, the two-component power-law fit yields $E(B-V) = 0.89$ mag.  

We see absorption in the \ion{Mg}{2} line, plotted in velocity space in the top right-hand panel of Figure \ref{fig:LRIS}.  The purple line is the continuum from our fit.  
We see two distinct absorption systems that are well fit by a double Gaussian, with full width at half maximum (FWHM) speeds of 4348 km s$^{-1}$ and 1838 km s$^{-1}$, and systemic outflow speeds of $-1407$ km s$^{-1}$ and $-350$ km s$^{-1}$.
A feature peaking at +3000 km s$^{-1}$ from the \ion{Mg}{2} position is unidentified and may be part of  the \ion{Mg}{2} line itself.

The other two panels of along the top of Figure \ref{fig:LRIS} display the same for \ion{C}{4} (middle) and Ly$\alpha$ (left), which both show a blueshifted absorption feature, also well fit by a Gaussian.  The feature at \ion{C}{4} is best fit by a single Gaussian with FWHM velocity width of 3060 km s$^{-1}$ and an outflow velocity of $-1576$ km s$^{-1}$. The absorption at Ly$\alpha$ is best fit by two Gaussians with FWHM velocity widths of 4184 km s$^{-1}$ 1306 km s$^{-1}$, outflowing at $-1013$ km s$^{-1}$ and $-515$ km s$^{-1}$. 
These features are indicative of outflowing gas.
 
\subsection{Black hole mass} \label{sec:mbh}

\begin{figure*}[ht!]
\includegraphics[scale=0.295]{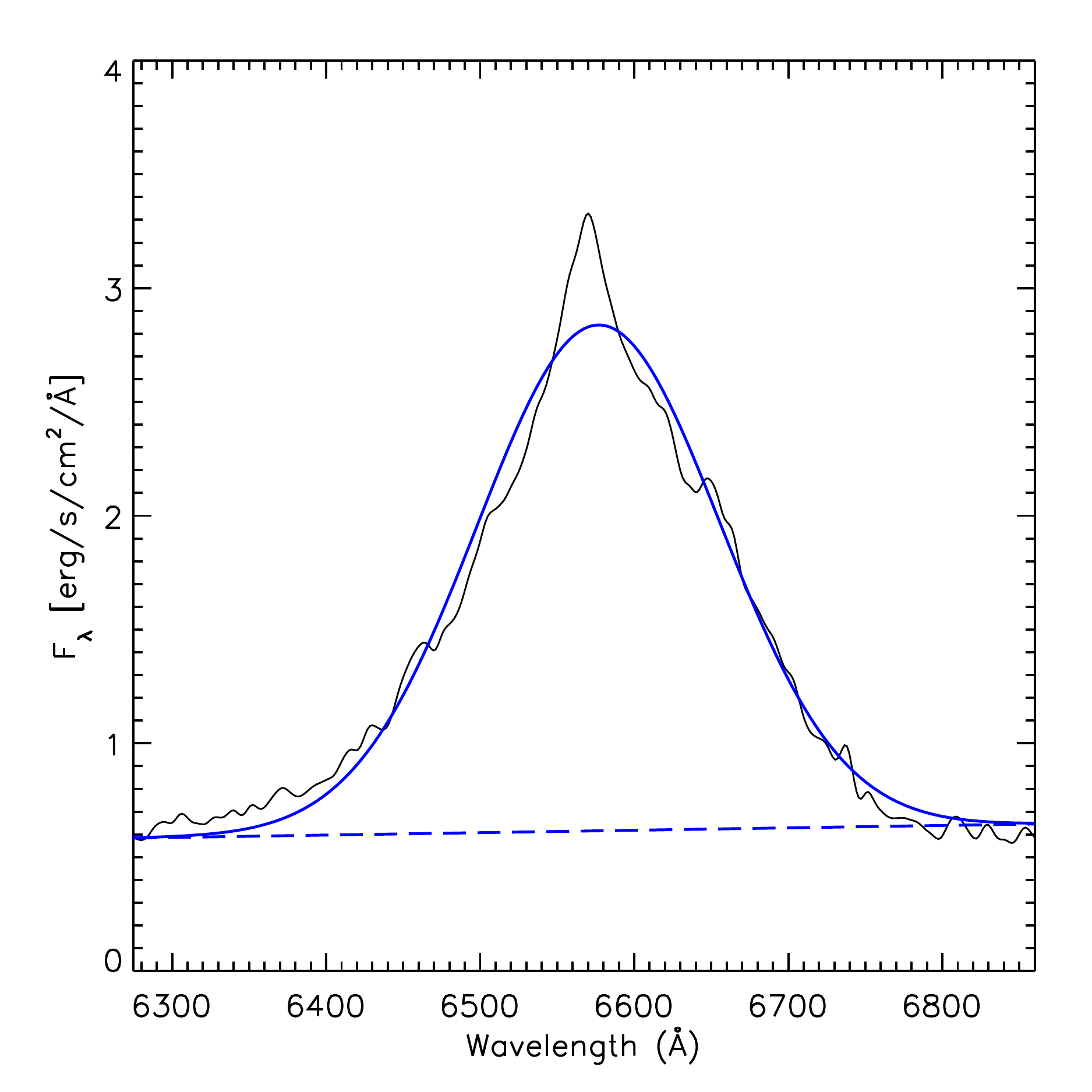}
\includegraphics[scale=0.32]{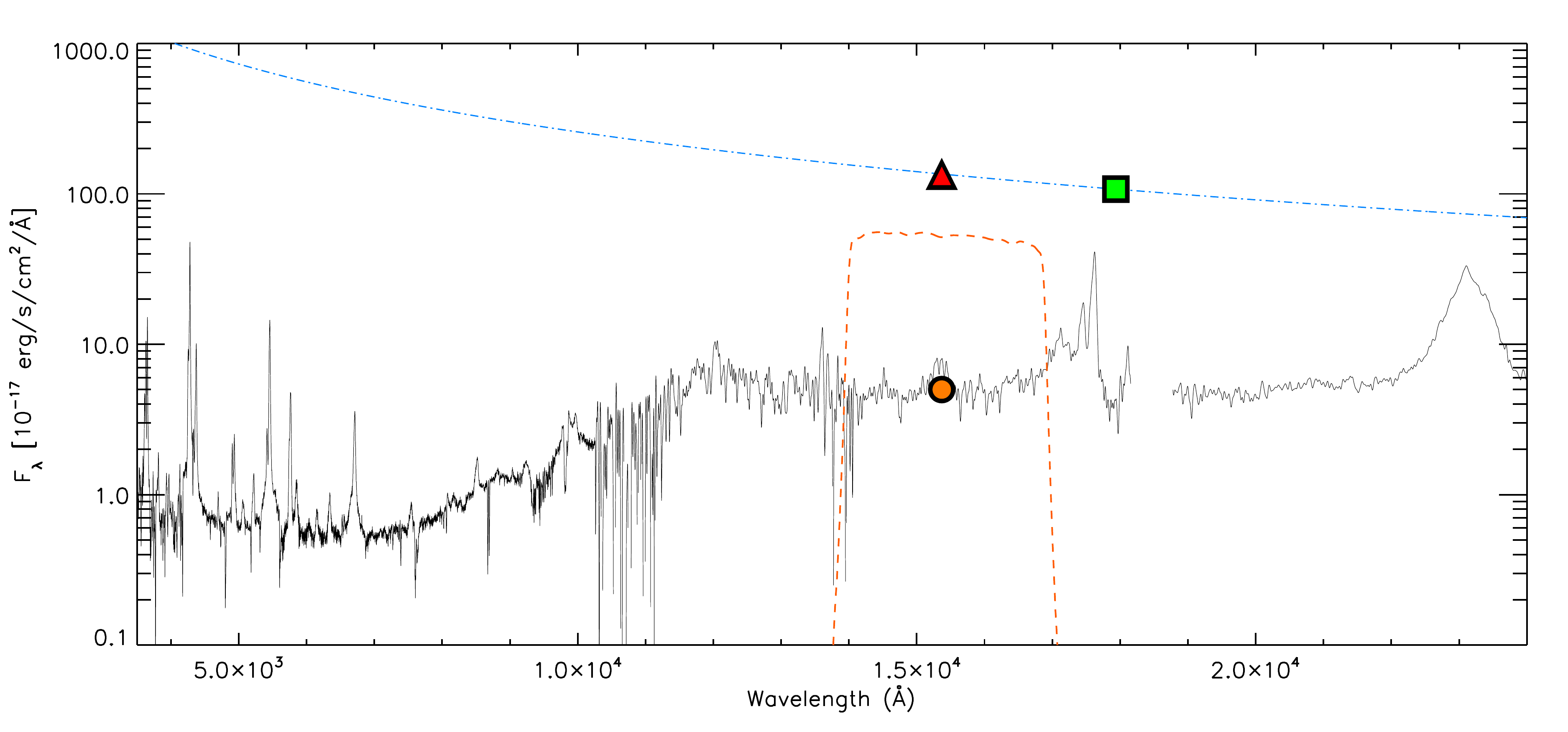}
\caption{{\em Left --} We show the H$\alpha$ line, seen in the near-infrared spectrum (black line), fit with a single Gaussian plus a linear continuum fit (solid blue line) to determine the FWHM needed to compute the black hole mass in Equation \ref{eqn:mbh}. The linear continuum is shown with a dashed blue line. {\em Right --} We scale our spectrum to the total F160W flux from the QSO (orange circle). We plot the de-reddened continuum from Equation \ref{eqn:cont} (cyan dot-dashed line) and shift from the observed flux through the F160W band to match the de-reddened continuum (red triangle) to the flux at 5100 \AA\ flux (green square). We use this value to find $L_{5100}$ and compute the black hole mass in Equation \ref{eqn:mbh}.
}\label{fig:mbh} 
\end{figure*}

Figure \ref{fig:mbh}, left, shows that the H$\alpha$ line in the near-infrared spectrum is well fit by a single Gaussian, with a FWHM in velocity space of $v_{\rm FWHM} = 8479$ km s$^{-1}$, which we use to estimate the black hole mass of W2M J1042+1641.  We combine the line width with an estimate of the QSO's intrinsic luminosity at 5100\AA\ and apply those values to the single-epoch virial black hole mass estimator ($M_{\rm BH,vir}$) following the formalism of \citet{Shen12},
\begin{equation}
\log \bigg(\frac{M_{\rm BH,vir}}{M_\odot} \bigg) = a + b \log \bigg(\frac{L_{5100}}{10^{44} \rm erg/s} \bigg) + c \log \bigg(\frac{v_{\rm FWHM}}{\rm km/s} \bigg),
\label{eqn:mbh}
\end{equation}
adopting the values $a=0.774$, $b=0.520$, $c=2.06$ for single-epoch measurements of FWHM$_{H\alpha}$ and $L_{5100}$, based on the calibration of \citet{Assef11}. 
We choose the H$\alpha$ line, because it is in a region of minimal reddening in our spectrum and the more-commonly-used H$\beta$ is strongly blended with [\ion{O}{3}].  The values derived from our procedure are listed in Table \ref{tab:mbh}.

To determine $L_{5100}$, we must account for reddening and magnification, as well as an absolute spectrophotometric calibration of our near-infrared spectrum. Our procedure is shown in the right-hand panel of Figure \ref{fig:mbh}. 
Because the magnification of the QSO is derived from the F160W image, de-magnifying the observed luminosity at this wavelength ($\lambda_{\rm F160W, rest} = 4370$\AA) will give the most internally consistent results.  
We first estimate the continuum flux, represented by Equation \ref{eqn:cont}, through the F160W band and scale it to the summed flux of the four QSO components listed in Table \ref{tab:photometry} (orange circle). 
We then find the intrinsic continuum at this wavelength by de-reddening the second term in Equation \ref{eqn:cont} and adding to it the first term (cyan dot-dash line).  
We scale the total observed flux from the four QSO components to match the de-reddened continuum and then shift their flux value at $\lambda_{\rm F160W, rest} = 4370$\AA\ (red triangle) to the de-reddened flux at $\lambda_{\rm rest} = 5100$\AA\ (green square).  
Finally, we divide the flux by the magnification factor of $\mu = 117$ (\S \ref{sec:magnif}) and compute a luminosity of $\log(L_{5100} ~{\rm [erg/s]) = 45.35}$.  These values yield $\log (M_{\rm BH,vir} ~[M_\odot]) = 9.56$.

To estimate $L/L_{Edd}$, we use a bolometric correction (BC) at 5100\AA\ of 9.5 from \citet{Richards06}. This gives $\log(L_{\rm bol} ~ {\rm [erg/s]}) = 46.33$ and a corresponding Eddington ratio, $\log(L/L_{Edd}) = -1.34$, which is a relatively low accretion rate compared with previously studied red quasars, most of which have $\log(L/L_{Edd}) \gtrsim -0.3$ \citep{Kim15}.

For comparison, recompute these physical parameters, using the observed {\em WISE} $W4$ ($22 \mu$m) instead of the F160W flux, as the rest wavelength of $W4$ corresponds to 6.28\um, which suffers minimal extinction compared with the F160W band and is dominated by AGN emission \citep{Stern14}.  
For example, an extinction of $E(B-V) = 0.7$ results in $A_{W4}=0.06$ mag (or 6\% of the flux) at 6.28\um, and thus avoids the need for a reddening correction.
In Section \ref{sec:substructlens}, we concluded that the flux anomaly seen in this object is due at least in part to substructure, and therefore the magnification factor relevant at this wavelength would be the same one observed in the {\it HST}/WFC3 F160W images ($\mu\sim117$)\footnote{If the anomaly was instead due entirely to microlensing, the magnification would be much closer to the value of $\mu\sim53$, as microlensing is expected to become insignificant at long wavelengths.}.
We scale the observed {\em WISE} $W4$ luminosity to the luminosity at 5100\AA\ using the spectral energy distribution (SED) from \citet{Richards06} for optically red QSOs. We de-magnify this flux by a factor of 117 and find $\log (L_{5100} ~{\rm [erg/s]}) = 44.37$, resulting in $\log (M_{\rm BH,vir} ~[M_\odot]) = 9.06$.

Using a BC of 7.5 from \citet{Richards06} based on the optically red QSO SED\footnote{We note that using the template for all SDSS QSOs (BC$\simeq7.9$) does not significantly affect the results.} for a frequency corresponding to $W4$ (22 $\mu$m) in the rest frame, $\log (\nu {\rm [Hz]}) = 13.68$, we compute $\log (L_{\rm bol} ~{\rm [erg/s]}) = 46.47$ -- consistent with the estimate based on the de-reddened and F160W flux. We find an Eddington ratio of $\log(L/L_{Edd}) = -0.68$, which is more consistent with the values found for other red quasars. 




\begin{deluxetable}{lcl}




\tablecaption{Black Hole Mass and Accretion Rate Estimates \label{tab:mbh}}

\tablenum{8}

\tablehead{\colhead{Parameter} & \colhead{Value} & \colhead{Unit} \\ 
\colhead{} & \colhead{} & \colhead{} } 

\startdata
QSO properties: & & \\
\hline
redshift, $z$        & 2.517         & \ldots \\
magnification, $\mu$ & 117           & \ldots \\
\hline
$H\alpha$ line fit: & & \\
\hline
FWHM$_{H\alpha}$    & $8340\pm330$  & km s$^{-1}$ \\
$F_{H\alpha}$       & $4.4\pm0.2 \times 10^{-14}$ & erg s$^{-1}$ cm$^{-2}$ \\
$\log(L_{H\alpha})$ &   44.09 & erg s$^{-1}$  \\
$\log(M_{BH})$     &    9.51 & $M_\odot$ \\
\hline
Continuum fit: & & \\
\hline
$F_{5100}$           & $1.07\times10^{-15}$ & erg s$^{-1}$ cm$^{-2}$ \AA$^{-1}$ \\
$\log(L_{5100})$     &   45.35    & erg s$^{-1}$ \\
$\log(M_{BH})$       &    9.56    & $M_\odot$ \\
$\log(L_{bol})$      &   46.33    & erg s$^{-1}$ \\
$\log(L/L_{Edd})$    & $-$1.34    & \ldots \\
\hline
WISE based: & & \\
\hline
$\log(L_{W4})$       &   45.60    & erg s$^{-1}$ \\
$\log(L_{5100})$     &    44.37   & erg s$^{-1}$ \\
$\log(M_{BH})$       &    9.06    & $M_\odot$ \\
$\log(L_{bol})$      &   46.47    & erg s$^{-1}$  \\
$\log(L/L_{Edd})$    & $-$0.68    & \ldots \\
\enddata


\tablecomments{All luminosities presented are de-magnified.}

\end{deluxetable}

\subsection{The relation between the black hole mass and the host galaxy luminosity and stellar mass}\label{sec:bhcorr}

Gravitational lensing provides a unique opportunity to study the co-evolution of SMBH host galaxies at high redshifts where, in the absence of lensing the AGN outshines the host making it near-impossible to cleanly separate the two components. 
Lensed QSOs offer the advantage of magnifying and stretching out the host galaxy while preserving surface brightness. At the same time, the AGN light, being a point source, appears in multiple distinct spots and is easier to separate and subtract than in the absence of lensing. Early work, using 31 lensed QSOs at high redshifts ($1<z<4.5$) to study their host galaxies and investigate co-evolution, was reported in \citet{Peng06}.  More recently, \citet{Ding17,Ding21} analyzed an additional eight lensed systems at $0.65 < z < 2.32$ including a direct measure of their stellar masses, $M_\star$, using SED modeling. Both studies observe an increase in the $M_{BH}/M_\star$ ratio toward earlier times, though the number of sources is statistically small above $z\sim 2.5$ and the \citet{Peng06} does not directly measure $M_\star$, but rather infers the ratio via the evolution of host luminosities, $L_R$, toward high redshift.  

We have determined the de-magnified magnitude of the host galaxy of W2M J1042+1641 in both {\em HST} bands (\S \ref{sec:coupledmodeling}), reported in the last column of Table \ref{tab:photometry}. When combined with the black hole mass that we estimated in Section \ref{sec:mbh}, we can include W2M J1042+1641 in these high redshift investigations of galaxy and SMBH co-evolution. 

Following \citet{Peng06}, we use the demagnified source magnitude in F160W to convert to absolute magnitude $R$-band, which is commonly used in the literature, as the K-correction is less dependent on the galaxy SED template.\footnote{We employ the \citet{Coleman80} spectral templates.} We note that while the de Vaucouleurs profile, used to fit the radial light profile of early-type galaxies, provides a better fit to the lensed host galaxy light, the Sbc template actually provides a better fit to the color (the galaxy is bluer in F125W than predicted from the E template). We find {\bf $\log(L_R/L_\odot) \sim 11.4$}  for the early-type template, and {\bf $\log(L_R/L_\odot) \sim11.2$} for the Sbc template.\footnote{To implement the K-correction, we use the \texttt{mag2mag} routine (see \S \ref{sec:lensz}).} 

\begin{figure*}
\begin{center}
\includegraphics[angle=0,scale=0.4]{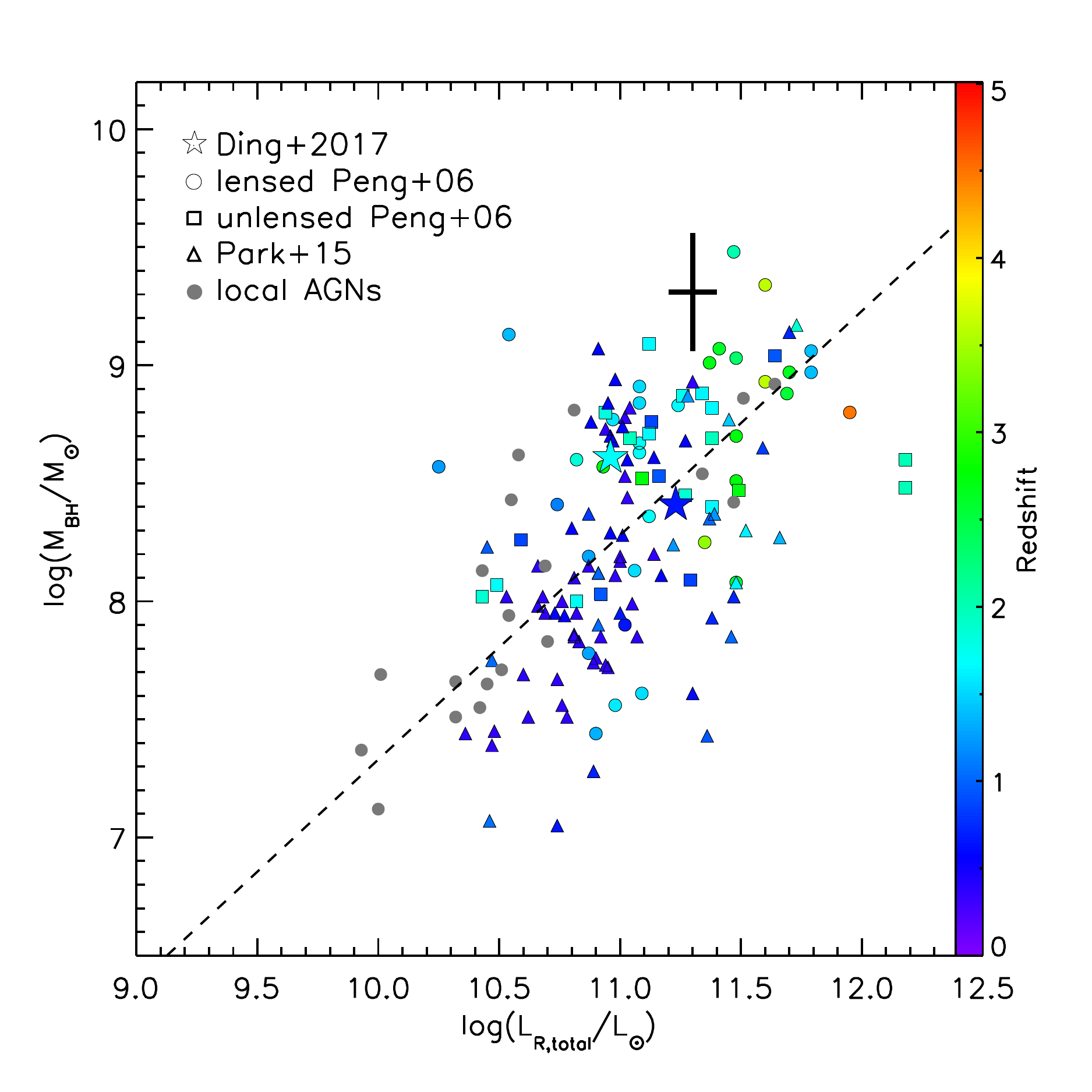}
\includegraphics[angle=0,scale=0.4]{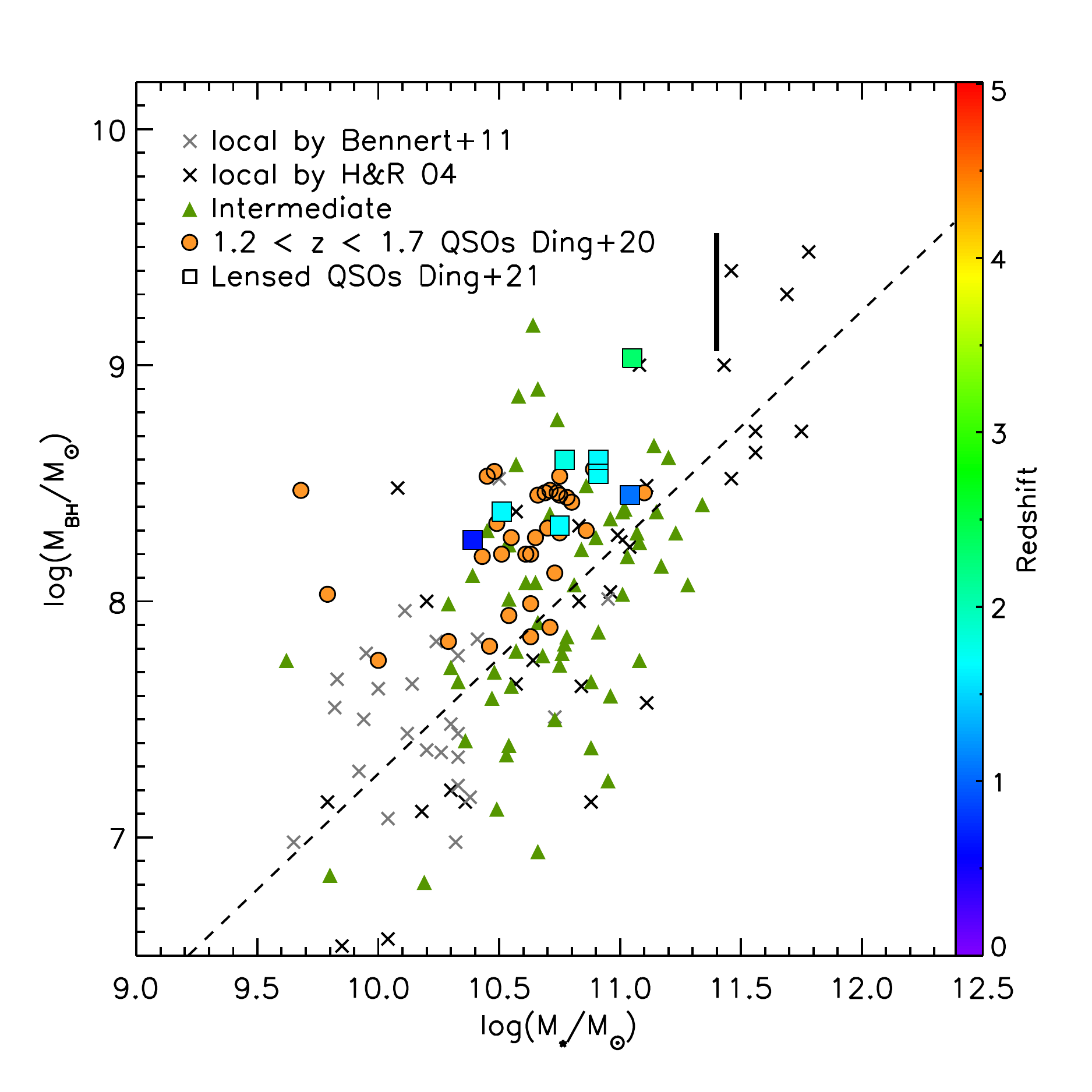}
\caption{{\em Left --} Observed black hole mass versus total $R$-band host galaxy luminosity using data from lensed and un-lensed AGN across a broad redshift range \citep[reproducing Figure 3 from][which has not been corrected for passive evolution]{Ding17}. 
{\em Right --} BH masses versus stellar mass, as shown in Figure 3 of \citet{Ding21}.  
Sources are color-coded by redshift and W2M J1042+1641 is represented by a thick black line spanning the possible values according to our calculations. The dashed black line is a fit to the local AGNs.  
In both figures, W2M J1042+1641 is above the local relation, and although it is the highest redshift object with a stellar mass estimate, its offset from the local relation is consistent with the next-highest redshift source, HE1104$-$1805 at $z=2.32$ \citep[Green circle;][]{Ding21}.  
}\label{fig:host-bh}
\end{center}
\end{figure*}

The left-hand panel of Figure \ref{fig:host-bh} plots $M_{BH}$ versus $L_R$ for samples of galaxies with these available measurements, following the structure of Figure 3 from \citet{Ding17} for ease of comparison.  Lensed and unlensed sources from \citet{Peng06} and \citet{Park15} are plotted with circles and squares, respectively, and are color coded by redshift as indicated by the colorbar. Two additional lensed QSOs at $z = 0.654$ and $z = 1.693$ from \citet{Ding17} are plotted with star symbols and local AGN from \citet{Bennert10} are shown with gray circles and the dashed black line represents the best fit to the local relation.  W2M J1042+1641 is plotted with a black cross representing the range of black hole masses as estimated in Section \ref{sec:mbh} and host luminosities from the K-correction described above. W2M J1042+1641 lies above the relation, in the region of the diagram consistent with other galaxies in the $z = 2-3$ range (green symbols).

To investigate the growth of the host galaxy's stellar mass, given only two photometric measurements of the host galaxy flux, we cannot use SED modeling as was done in \citet{Ding21}.  However, at the redshift of W2M J1042+1641, the $F125W$ and $F160W$ effective wavelengths shift near perfectly to the SDSS $u$ and $g$ passbands, respectively.  Therefore, we can use the $F125W-F160W$ color as rest-frame $u-g$ and estimate the galaxy's stellar mass using mass-to-light ratios from a single color as determined by \citet{Bell03} (Table 7). Using the rest-frame $g$-band luminosity, we estimate $\log(M_\star/M_\odot) = 11.4$ with the model that includes GX as a perturber\footnote{$\log(M_\star/M_\odot) = 11.5$ is estimated for the model without a perturber}. 

We caution that this stellar mass is based on only two photometric measurements, and is further subject to the uncertainty propagated from the magnification factor, but is the best that can be estimated with the current data. An estimate of the stellar mass in W2M J1042+1641 was also conducted by \citet{Matsuoka18}, which found $\log(M_\star/M_\odot) = 13.55$ using an SED fit \citep[with the Code Investigating GALaxy Emission, CIGALE;][]{Noll09} to the observed photometry which includes both the AGN and host galaxy light. They then de-magnify the SED by $\mu=122$\footnote{This was an initial estimate for the magnification, reported in \citet{Glikman18}, but which is superseded by the more thorough analysis presented in this work.} to arrive at $\log(M_\star/M_\odot) = 11.46$.
Despite the coincidental agreement between our values, we caution that the SED fitting method in \citet{Matsuoka18} contains several inaccuracies in this approach. First, the CIGALE SED fitting code does not include a component for a reddened AGN and only allows for either an unreddened AGN (Type 1) or a completely obscured AGN (Type 2), neither of which fits the shape of the moderately reddened, yet AGN-dominated continuum of this red QSO. This leads CIGALE to attribute the rest-frame UV-optical light to a heavily reddened host galaxy, which will naturally over-estimate $M_\star$.  Furthermore, the CIGALE-reported $M_\star$ is based on an SED whose galaxy component was de-magnified by the same factor as the AGN. However, our analysis shows that the the (extended) host galaxy is only magnified by a factor of 15 (computed with \texttt{hostlens} for the SIE+$\gamma$+GX model), which is off by a factor of $\sim8$ from the AGN's magnification (from equation \ref{eq:mu}). It appears that the overestimate of the stellar mass from the SED fitting and the over-correction of the host galaxy's de-magnification conspire to yield the same value that we derive here. Depending on the value for $M_{BH}$ used, from Table \ref{tab:mbh}, we find $\log{M_{BH}/M_\star} = -2.3$ to $-1.8$.

The right-hand panel of Figure \ref{fig:host-bh} plots $M_{BH}$ versus $M_\star$, following the structure of Figure 3 from \citet{Ding21} for ease of comparison. This plot has fewer high redshift sources due to the necessity of multiple bands for estimating $M_\star$.  Here, the local AGNs are represented by gray and black $\times$'s for the samples of \citet{Bennert11a} and \citet{HR04}, respectively. Intermediate-redshift sources spanning $0.34 < z < 1.90$ are plotted with green triangles \citep{Bennert11b,Cisternas11,Schramm13}. Intermediate redshift, unlensed QSOs from \citet{Ding20} are shown in orange circles, and the sample of lensed QSOs analyzed in \citet{Ding17,Ding21} are plotted with squares, and colored according to their redshift coded by the colorbar. 
W2M J1042+1641 appears as a vertical black line, spanning the range of possible BH masses at the stellar mass of $\log(M_\star/M_\odot) = 11.4$ that we derived. W2M J1042+1641 is the highest redshift source among the objects in this Figure, and is shifted by a similar amount from the local relation (black dashed line) as the next highest redshift source, HE1104$-$1805 at $z=2.32$.

\subsection{Magnification and population analysis}\label{sec:population}

\begin{figure}[ht!]
\plotone{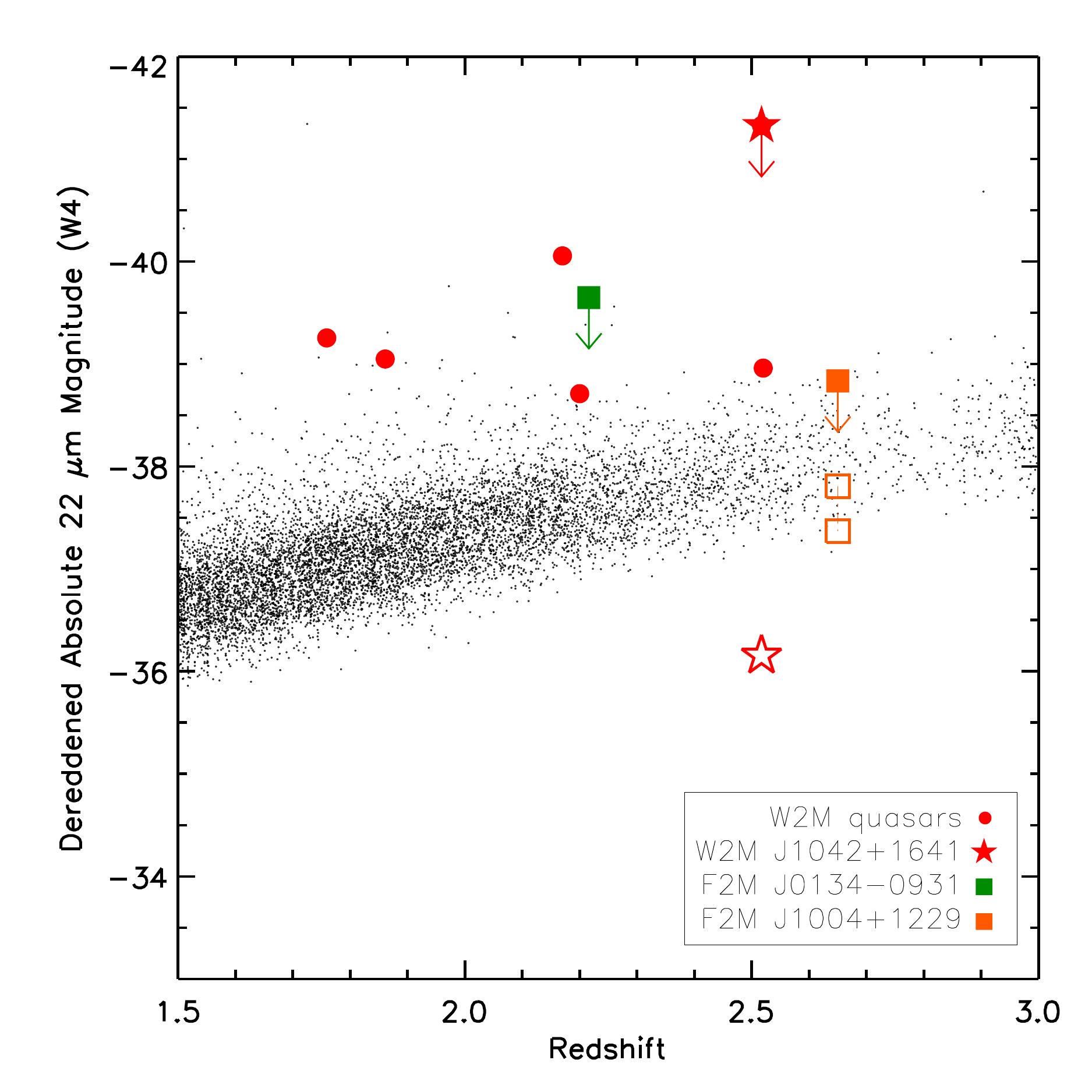}
\caption{Absolute magnitude at 22\um~ {\bf in the observed frame}, from {\em WISE}, for the sample of red QSOs in the W2M survey.  The observed luminosity of W2M~J1042+1641 is indicated by a star symbol, with a downward-facing arrow pointing toward its de-magnified intrinsic luminosity (using $\mu = 117$). This figure indicates that W2M~J1042+1641 is a reddened QSO drawn from a population of sources that is not accessible to the wide-area QSO samples currently being studied. We show, for comparison, the observed and magnification-corrected luminosities of the other high-redshift lensed F2M red quasars discussed in the text, using only an upper limit for F2M~J0134-0931 (see text) and the luminosity range for F2M~J1004+1229 corresponding to magnification $3.15^{+0.71}_{-0.55}$. } 
\label{fig:lumz}
\end{figure}

Figure \ref{fig:lumz} shows W2M~J1042+1641 (red filled star) on a WISE $W4$ luminosity versus redshift diagram. The filled red circles are the other high-redshift W2M QSOs found in our study, which is spectroscopically complete \citep{Glikman22}. We compare them with $\sim20,000$ QSOs from SDSS with matches in the UKIDSS and WISE catalogs with spectroscopic redshifts from \citet{Peth11} (black dots). 
The de-magnified position of  W2M~J1042+1641 is shown as an open star, below the lower envelope of SDSS QSOs detected which represents the WISE $W4$ detection limit. We see that this red QSO exists among a population whose luminosity is too low to be discovered in the wide-field surveys that yield the vast majority of QSOs in the literature.  

We also plot in Figure \ref{fig:lumz} the positions of the two other lensed F2M red quasars: F2M J0134$-$0931 \citep[green square;][]{Gregg02} and F2M J1004+1229 \citep[orange square;][]{Lacy02}. We modeled F2M J1004+1229 by fitting SIE$+\gamma$ models to the relative astrometry reported in CASTLES\footnote{CfA-Arizona Space Telescope LEns Survey, C.S. Kochanek, E.E. Falco, C. Impey, J. Lehar, B. McLeod, H.-W. Rix, \url{https://www.cfa.harvard.edu/castles/}}, in order to determine its magnification factor. We do not compute a magnification factor for F2M J0134$-$0931, as this system has a complex lensing configuration \citep{Keeton03}, whose modeling is beyond the scope of this paper.
Their de-magnified positions are shown with corresponding open symbols and they, too, lie at the edge of or below the faintest luminosities accessible to SDSS, UKIDSS, and {\em WISE}. 

The W2M survey finds 7 high redshift ($z>1.7$) QSOs, including the lensed quasar F2M~J1004+1229, which we recover from our previous FIRST-selected sample. That means that the lensing fraction of this complete, flux-limited sample is 2/7 = 28\% -- three orders of magnitude higher than the lensing fraction for luminous unobscured QSOs in typical surveys \citep{OM10}. 
The lensing fraction of the F2M survey (Section \ref{sec:intro}) is also large, at $\sim2$\%, but smaller than that of the W2M survey. This is likely due to the F2M survey being about a magnitude deeper than W2M.

We note that the previously-discovered lensed red quasars are all radio sources. Some of them are intrinsically reddened, while others are likely reddened by the lensing galaxy itself.  
\citet{Malhotra97} showed that lensed quasars found in radio-selected samples have redder colors, implying that radio selection finds dusty systems that may be lost in optical QSO samples that often impose a blue color cut.  
Our surveying of red QSOs in shallow, wide-field surveys (i.e., W2M) is beginning to remedy this incompleteness by recovering radio-quiet reddened lensed QSOs. In addition, shallow flux-limited surveys benefit from increased magnification bias, making these lenses easier to find.  

The de-magnifed luminosity density computed from the $W4$ flux density of W2M~J1042+1641 ($\lambda_{\rm rest} = 6.2$ \um) is $\log(L) = 32.46$ in units of erg s$^{-1}$ Hz$^{-1}$ for magnification = 117. 
We interpolate between the {\em WISE} bands and find the de-magnified $\log(L_{5\mu{\rm m}}) = 32.29$. 
The 5\um\ luminosity function at $z=2.5$ for red QSOs in deep {\em Spitzer} fields is $\log(L_0^*) = 31.92$ (Tab.~8 and Fig.~19 of \citealt{Glikman18}).
Thus, when demagnified, W2M~J1042+1641, is near the knee but on the bright-end side of the red QSO luminosity function (QLF) derived in \citet{Glikman18}.  
The sources that enabled a determination of this QLF were derived from deep {\em Spitzer} fields that covered relatively small areas.  
Although we cannot use this QSO to investigate the faint end of the red QLF, finding such a highly magnified red QSO offers a unique opportunity to study a population of QSOs whose luminosity is intrinsically lower than the QSOs found in wide-field surveys such as SDSS, which make up the vast majority of known QSOs.  

\section{Conclusions} \label{sec:conclusions}

We have discovered a quadruply lensed radio-quiet QSO at $z=2.517$ identified through {\em HST} imaging of dust-reddened QSOs selected by their {\em WISE} colors. Using optical and near-infrared spectroscopy, we determine that the QSO is reddened by $E(B-V) \simeq 0.7 - 0.9$ from dust intrinsic to the QSO's environment, as opposed to dust in the lensing galaxy. Our lensing analysis finds a magnification factor of $\sim 52$ for the best-fit model, but boosted to $\sim 117$ due to strong flux anomaly. Using photometric data from near-IR to radio, we conclude that substructure is the most likely cause for the anomaly.

We estimate the QSO's black hole mass to be in the range $\log (M_{\rm BH} ~[M_\odot]) = 9.06 - 9.56$, depending on how the unreddened continuum luminosity is computed. 
The QSO's rest-frame infrared luminosity is $\log (L_{5\mu{\rm m}}~[{\rm erg~s^{-1}~Hz^{-1}}]) = 32.29$, which is near the knee of the QLF, representing more typical quasar luminosities that are difficult to access at high redshifts. 
The QSO's Eddington ratio could be as high as $L/L_{Edd} = 0.21$, if the bolometric luminosity is estimated from the 22\um\ flux, but could be as low as $L/L_{Edd} = 0.05$ if the {\em HST} F160W band is used.  The former would be consistent with accretion rates seen for more luminous red quasar samples. 
These characteristics, in addition to evidence for outflowing gas seen in absorption in the QSO's spectrum, point to a system in a transtional phase following a major merger, as is seen in the hosts of more luminous red quasars. 

In the future, on a long timescale, monitoring observations would be required to separate microlensing and intrinsic flux variations. Deeper radio observations to robustly measure the flux of all four QSO images, or observations probing narrow line emission from the source, too spatially extended to be affected by microlensing \citep{Nierenberg14,Nierenberg17}, would be required to further separate the effects of microlensing and substructure, and provide a more robust total magnification factor. 
In the shorter term, a dedicated effort to completing the pixelated modeling, described in Appendix \ref{sec:host}, with both filters and both visits will improve our understanding of the QSO host galaxy and its apparent companion (source X). In addition, a more thorough analysis, adding in the two WFC3/UVIS bands to the WFC3/IR bands will constrain the colors and SED of the host galaxy to better probe co-evolution of QSOs and their hosts at the height of cosmic noon.  

\acknowledgments

We thank the anonymous referee for their careful reading of the manuscript and thoughtful suggestions, which have significantly improved the quality of our final work. 

EG acknowledges the generous support of the Cottrell Scholar Award through the Research Corporation for Science Advancement. 
E.G. is grateful to the Mittelman Family Foundation for their generous support. 
This work was performed in part at Aspen Center for Physics, which is supported by National Science Foundation grant PHY-1607611.  

CER thanks Paul Schechter for useful discussions. 
CS is grateful for financial support from the National Research Council of Science and Technology, Korea (EU-16-001) and from the Italian Ministry of University and Research - Project Proposal CIR01\_00010.
SGD and MJG acknowledge a partial support from the NSF grants AST-1413600 and AST-1518308, as well as the NASA grant 16-ADAP16-0232.

This work is based on GO observations made with the NASA/ESA Hubble Space Telescope from the Mikulski Archive for Space Telescopes (MAST), which is operated by the Association of Universities for Research in Astronomy, Inc., under NASA contract NAS5-26555. These observations are associated with program \#14706.

This paper makes use of the following ALMA data: ADS/JAO.ALMA\#2019.1.00964.S. ALMA is a partnership of ESO (representing its member states), NSF (USA) and NINS (Japan), together with NRC (Canada), MOST and ASIAA (Taiwan), and KASI (Republic of Korea), in cooperation with the Republic of Chile. The Joint ALMA Observatory is operated by ESO, AUI/NRAO and NAOJ.

We thank the staff at the Keck observatory, where some of the data presented here were obtained. The authors recognize and acknowledge the very significant cultural role and reverence that the summit of Maunakea has always had within the indigenous Hawaiian community. We are most fortunate to have the opportunity to conduct observations from this mountain.

The National Radio Astronomy Observatory is a facility of the National Science Foundation operated under cooperative agreement by Associated Universities, Inc.

This work makes use of the following Python packages: Astropy \citep{astropy13}, Numpy \citep{walt11}, Scipy \citep{virtanen20} and Matplotlib \citep{hunter07}.

\vspace{5mm}
\facilities{HST(WFC3/IR), Keck(LRIS), IRTF(SpeX), LBT(MODS1B), NRAO(VLA), ALMA}
\software{Spextool \citep{Cushing04}, IRAF (Tody 1986, Tody 1993), DrizzlePac (Hack et al. 2012), glafic \citep{Oguri10}, galfit \citep{Peng02}}

\appendix

\section{SIE+$\gamma$ model}  \label{SIEgamma}
\begin{figure}[ht!]
\begin{center}
\includegraphics[angle=0,scale=0.6]{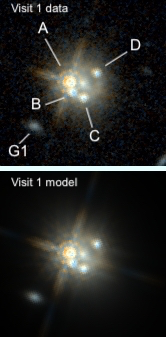}
\includegraphics[angle=0,scale=0.6]{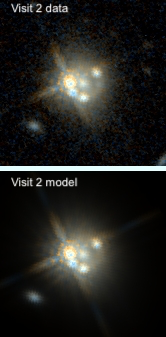}
\caption{{\em HST} WFC3/IR F125W and F160W color-combined images of W2M~J1042+1641 over two visits, along with output from a morphological analysis with {\tt hostlens}, as in Figure \ref{fig:hst1}, but with a ``free SIE+$\gamma$'' model.
}\label{fig:hst1_nopert}
\end{center}
\end{figure}

\begin{figure}[ht!]
\begin{center}
\includegraphics[angle=0,scale=1]{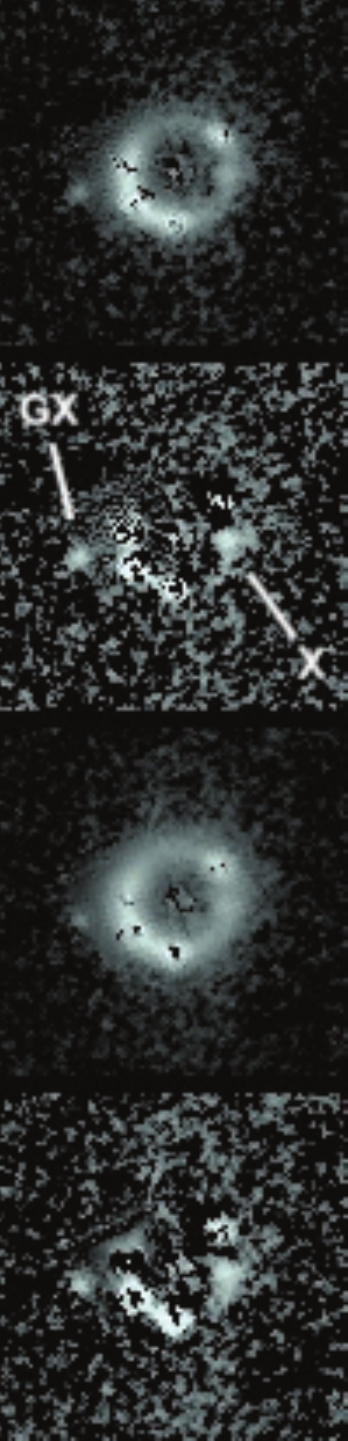}
\includegraphics[angle=0,scale=1]{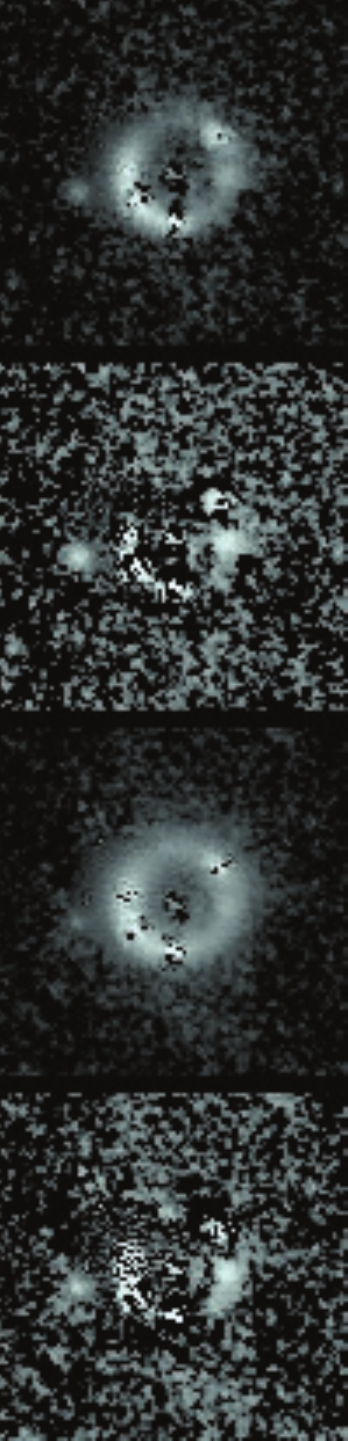}
\caption{{\em HST} Residual images of W2M~J1042+1641 over two visits, after subtracting the best-fit morphological models determined by {\tt hostlens} shown in the bottom panel of Figure \ref{fig:hst1_nopert}, as in Figure \ref{fig:hst1_resid}, but with a ``free SIE+$\gamma$'' model.}\label{fig:hst1_resid_nopert}
\end{center}
\end{figure}

We show in Figures \ref{fig:hst1_nopert} and \ref{fig:hst1_resid_nopert} the results of our fitting using the SIE+$\gamma$ model discussed in \S \ref{sec:lensing} and Table \ref{tab:lens}. This model does not include object GX as a perturber. Despite this, the difference between images is barely distinguishable by eye and our reasoning for adopting the model with GX as a perturber (SIE+$\gamma$+GX) is laid out in \S \ref{sec:lensing} and \S \ref{sec:discussion}.

\section{Details of the microlensing calculation} \label{microlensdetails}

We follow the simulation technique in \citet{chen18}, which itself is built on the approach in \citet{kochanek04} to compute microlens tracks. The technique is implemented in the code \texttt{GPU-D}\footnote{\url{https://gerlumph.swin.edu.au/software/\#gpud}} \citep{vernardos14}, further modified by \citet{chan20} to account for the initial mass function (IMF). For the population of lenses we assume a Salpeter \citep{salpeter55} IMF and a mean solar mass of $0.3 M_\sun$. 
To characterize the QSO accretion disk, we consider the two values of the black hole mass, corresponding bolometric luminosities and Eddington ratios computed in \S\ref{sec:mbh} (cases 1 and 2 in Figure \ref{fig:microlensing}), and we compute the radiation efficiency $\eta$ by combining equations 11 and 13 from \citet{davis11}. For the size of the accretion disk, $R_0$, in the observed F160W filter we use Equation 2 from \citet{morgan10}, and assume inclination $i=60$ deg and PA$=90$ deg. To compute the microlensing time scale represented in the horizontal axis of the left plot in Figure \ref{fig:microlensing}, we follow the methodology in \citet{kochanek04}, and compute a typical microlens relative speed of $\sim340$ km/s. To compute the microlensing probability we used the observed magnification of image A and divided it by its predicted SIE$+\gamma$+GX model magnification, anchored by C/D in \S\ref{sec:magnif}. The magnification excess is $4.9\pm1.0$, and we measure the the fraction of pixels from the microlensing maps that satisfy these constraints.

\section{The QSO host galaxy, structure along the Einstein ring and pixelated modeling}\label{sec:host}

As mentioned in \S \ref{sec:morphmodeling}, we fitted the source QSO host galaxy with \texttt{hostlens}, using a circular de Vaucouleurs profile. 
The de-lensed magnitude of the source is given in Table \ref{tab:photometry}. 
As noted in \citet{Marshall07}, a dominant source of uncertainty on the magnitude of the source is due to the distribution of power law slope values in the radial profile of lenses (we have assumed isothermal profiles in \ref{sec:lensing}).
Following the appendix of \citet{Marshall07}, this uncertainty amounts to $\sigma_{mag}\sim0.26$ mag; the same uncertainty also affects the inferred effective radius $r_e$ as $\sigma_{r_e}/r_e\sim0.12$.
The effective radii inferred by \texttt{hostlens} are $0.15\arcsec$ or 1.2 kpc at the QSO redshift (F125W), $0.34\arcsec$ or 2.7 kpc (F160W) for the SIE$+\gamma$+GX model, and $0.16\arcsec$ or 1.3 kpc (F125W), $0.50\arcsec$ or 4.0 kpc (F160W) for the SIE$+\gamma$ model. Note that we find $r_e$ to be significantly larger in F160W.

The QSO host galaxy, lensed into an Einstein ring, does not appear to be smooth, but instead shows at least one blueish peak present in both visits and both filters, South of image D (denoted by ``X'' in Figure \ref{fig:hst1_resid}). 
This object shows what appears to be a tail following the curvature of the Einstein ring, which makes it unlikely to be a projected structure or one that is associated with the lensing galaxy. In fact, its location on top of the Einstein ring {\bf suggests} that X is most likely to be a source located at the same redshift with the QSO host. 
Thus, X must be an extended structure, part of the QSO host or of its close environment. The position of X on top of the Einstein ring implies that it must produce counterimage(s) (mirror-like image(s) located on the other side of the Einstein ring, with respect to the lens). Our approach in \S \ref{sec:coupledmodeling} is not designed to account for these, and they would be treated as noise around the bright QSO images during the PSF reconstruction. To identify such counterimages, we conduct an exploratory analysis where we model the first visit in the F125W filter by reconstructing the source on a grid of pixels, with the code \texttt{glee} \citep{Suyu10,Suyu12}. The advantage of using a pixellated grid is that it does not need to assume a particular morphology of the source object. 

\begin{figure*}
\epsscale{1}
\plotone{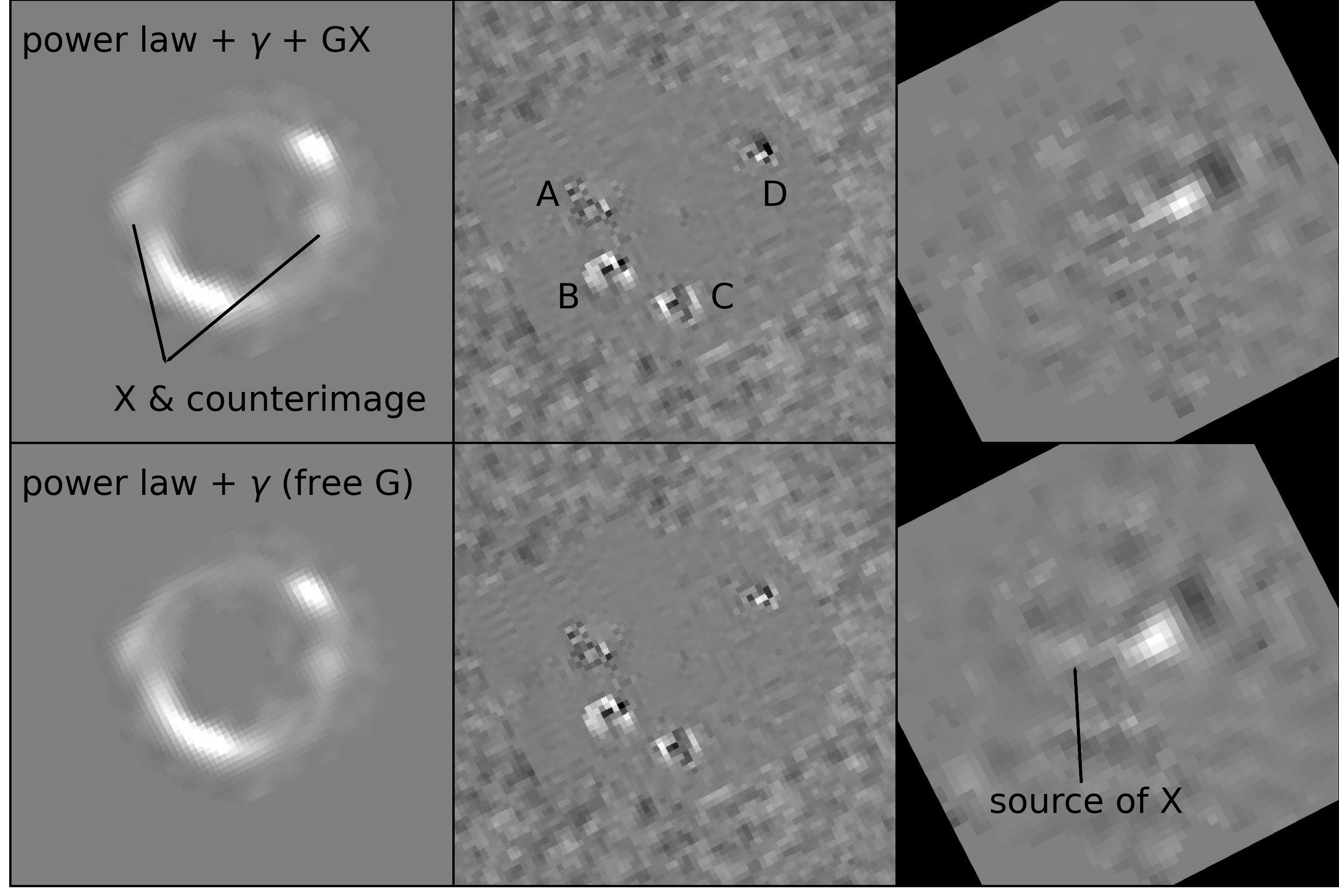}
\caption{Pixel-based modeling of F125W visit 1 with \texttt{glee} \citep{Suyu10,Suyu12}. A power law + $\gamma$+GX mass model was used for the upper plots, and a power law + $\gamma$ model with free lens position was used for the lower plots. From left to right: the best-fit model of the extended arcs, the residuals after subtracting the best-fit model (the gray scale is linear, from -10 $\sigma$ to 10 $\sigma$), and the source reconstruction on a $50\times50$ pixel grid. The pixel scales in the source plane averaged between the two axes are $0.030\arcsec$ (upper plot) and $0.024\arcsec$ (lower plot). North is up and East is to the left. Note in the arc the luminous blob, X, south of image D, and its fainter counter-image situated on the diametrically opposed part of the arc. The QSO host appears elongated with a possible tidal tail extending toward the less luminous substructure on the left, responsible for the luminous blob seen in the arc. 
}\label{fig:extendsrcmodel} 
\end{figure*}

Since \texttt{glee} uses a power-law rather than SIE for the mass model, we perform the modeling with a power law + $\gamma$+GX mass model with a slope of 2.22, and a power law + $\gamma$ model with free lens position with a slope of 1.90.
We show the structure of the pixel-based Einstein ring model, the residuals of the model and the reconstructed source structure in the source plane in Figure \ref{fig:extendsrcmodel}, for lensing models w/ and w/o GX. 
The Einstein ring model shows a clear extended counterimage of X to the East, and the reconstructed source appears elongated (Figure \ref{fig:extendsrcmodel}, top row). Since we were able to fit the Einstein ring with a circular analytical source profile in \S \ref{sec:coupledmodeling}, we interpret the source elongation to be due to the blending (given the reconstructed source resolution) between the QSO host galaxy and the source responsible for X and its counterimage. 
In the bottom row of Figure \ref{fig:extendsrcmodel}, rightmost panel, the source of X itself may even be resolved and possibly connected to the host galaxy by a tidal tail.
The existence of this extended source is in agreement with the observation that most red quasars are hosted by major mergers \citep{Urrutia08,Glikman15}. However, in this case, the reconstructed source resolution does not allow us to unambiguously reject the possibility that X could be a minor merger or even a star-burst or star-forming region in the host, caused by a merger.

\bibliography{w2m_lens.bbl}

\end{document}